\documentclass[onecolumn,tighten]{aastex631}

\shorttitle{Discovering the Mass-Scaled Damping Timescale}
\shortauthors{Zhang et al.}
\graphicspath{{figures/}}
\usepackage{amsmath}
\usepackage{multirow}
\usepackage{tipa}
\begin{document}
\title{Discovering the Mass-Scaled Damping Timescale from Microquasars to Blazars}
\author[0000-0003-3392-320X]{Haoyang Zhang}
\affiliation{Key Laboratory of Astroparticle Physics of Yunnan Province, Department of Astronomy, Yunnan University, Kunming 650091, China}
\author{Shenbang Yang}
\affiliation{Faculty of Science, Kunming University of Science and Technology, Kunming 650500, China}
\author[0000-0001-7908-4996]{Benzhong Dai}
\correspondingauthor{Benzhong Dai}{\email{bzhdai@ynu.edu.cn}}
\affiliation{Key Laboratory of Astroparticle Physics of Yunnan Province, Department of Astronomy, Yunnan University, Kunming 650091, China}

\begin{abstract}
Studying the variability of the accretion disks of black holes and jets is important to identify their internal physical processes. In this letter, we obtain the characteristic damping timescale of 34 blazars and seven microquasars from the Fermi-Large Area Telescope and the XMM-Newton X-ray telescope, respectively. We found that the mass-scaled characteristic timescales, ranging from the microquasars of stellar-mass black holes to the blazars of supermassive black holes, exhibited a linear relationship with a slope of $\sim$ 0.57. Given the fact the damping timescales of the $\gamma$-ray in the blazars are associated with the jet, we propose that the timescales of the X-ray in these microquasars are also related with the jet. The mass-scaled damping timescale that we found was consistent with the radiation of the optical accretion disk. This can be attributed to the viscous timescale at the ultraviolet-emitting radii of the disk, which can affect the jet. Our study provides a new perspective on the origin of the region of radiation and the possible disk--jet connection based on time-domain analysis.
\end{abstract}
\keywords{galaxies: jets --- Galaxy: disk --- (galaxies:) BL Lacertae objects: general --- methods: data analysis --- methods: statistical}
\section{Introduction} \label{sec:intro}
Some black hole accretion systems, from stellar-mass black hole (SBH) to supermassive black hole (SMBH) systems, i.e. microquasars \citep{1999ARA&A..37..409M} to blazars \citep{1995PASP..107..803U,2017A&ARv..25....2P,2019ARA&A..57..467B}, have been observed to have jet structures. In these systems, relativistic jets may be launched via the extraction of energy from the rotating black hole and/or from the angular momentum of the accretion flow in the presence of a large-scale magnetic field \citep{1977MNRAS.179..433B,1982MNRAS.199..883B}. Further theoretical frameworks based on these mechanisms have been proposed \citep{1995PhRvD..51.5387W,2004ApJ...611..952W,2014Natur.510..126Z,2018PPCF...60a4006B}. Although the exact physical origin of such jets is still unclear, these theoretical studies indicate that accretion disks play an important role.

Observed phenomena, both direct and indirect, indicate a connection between the accretion disk and the jet. The radio jet kinetic power has been shown to correlate significantly with the accretion disk luminosity in blazars, which may be evidence for disk--jet coupling \citep{2003ApJ...593..667M,2004ApJ...615L...9W,2017ApJ...840...46I}. The X-ray spectra of microquasars usually change between the hard and soft states \citep{2000A&A...355..271B,2002ApJ...578..357Z,2007A&ARv..15....1D}. Meanwhile, radio jet emission has been detected for many microquasars and a radio--X-ray correlation has been observed during the hard state \citep{2001A&A...372L..25M,2003A&A...400.1007C,2005ApJ...635.1203M,2006AA...447..245W,2020ApJ...894L..18Z}. The X-ray jet structure of the microquasar XTE J1550-564 has even been observed in the hard state \citep{2002Sci...298..196C}. This phenomenon may be a process of accretion flow energy injection into the jet \citep{2003ApJ...593..184L} or an expanded corona providing a base for jet emission \citep{2021ApJ...910L...3W}; indeed, the corona is closely related to accretion disks \citep{2014ARA&A..52..529Y}. This relationship, called the "fundamental plane of black hole activity", extends to SMBH objects \citep{2003MNRAS.345.1057M,2006A&A...456..439K,2016MNRAS.455.2551N}, which reveals the connection between the accretion disk and the jet in a wide range of black hole scales.

Accretion disks and jets typically have significant variability across the entire electromagnetic spectrum \citep{1989A&A...226...59K,1996ASPC..110..391U,1998ApJ...504..671K,2001AJ....122.2901D,2009MNRAS.392.1181D,2015ApJS..218...18D,2021ApJS..253...10F,2022ApJS..260...47C,2023Natur.621..271T}. An effective method to study such variability is to extract the power spectral density (PSD) from an observed light curve (LC). The PSDs of disks and jets are generally considered to be a bending power-law (BPL) \footnote{I.e., a power-law with a slope of $\sim$2 that transitions to white noise at the characteristic damping timescale, i.e. $\tau_{damping}=1/(2\pi f_{bend})$.}. The characteristic damping timescale ($\tau_{damping}$; hereafter referred to simply as the characteristic timescale) of an accretion disk may be related to some important physical processes, including thermal (related to restoring thermal equilibrium), dynamical (related to orbital motion), and viscous (related to mass flow diffusion) processes \citep{1999MNRAS.303..148C,2006ASPC..360..265C,2021ApJ...907...96S}. \citet{2006Natur.444..730M} discovered a scaling relationship ($\tau_{damping} \propto M_{BH}$) in the X-ray emissions of non-jetted SMBH systems for the first time. Subsequently, \citet{2015SciA....1E0686S} and \citet{2021Sci...373..789B} found a scaling relationship between non-jetted SBHs to SMBHs in optical band, which implies that black hole accretion systems with different masses have the same physical nature. \citet{2019MNRAS.486.1672M} suggested that the characteristic timescales of the accretion disk and the jet may be different. However, increasingly more results have indicated that the characteristic timescale of jet radiation is related to, or of the same order of magnitude as, the accretion disk in long-term LCs \citep{2012ApJ...760...51R,2019ApJ...885...12R,2022ApJ...930..157Z,2023ApJ...944..103Z,2024MNRAS.527.2672S}. If the disk--jet connection is significant enough, jet systems across a range of scales should also follow the scaling relationship. However, observations of jetted black hole systems with intermediate-mass black holes are currently lacking, which means finding this pattern across the jetted systems of all black hole masses is very difficult.

The $\gamma$-ray radiation of blazars is thought to be produced by inverse Compton scattering inside the jet \citep{2017A&ARv..25....2P,2019ARA&A..57..467B}. The origins of the X-ray emissions of microquasars remain uncertain. In general, the soft state X-ray emissions from microquasars are thought to come mainly from the accretion disk, which accounts for more than 75\% of the total X-ray emissions. In the hard state, the radiation of disk accounts for less than 20\% of the overall X-ray emissions \citep{2006ARAA..44...49R}. Some researchers have proposed that the X-ray emissions in the hard state may also originate from an advection-dominated accretion flow (ADAF), (\citealt{1997ApJ...482..448N,2022Sci...378..650K}). Therefore, X-ray emissions in the hard state can be used to investigate the jet physics of microquasars.

Gaussian processes (GPs) are widely used as a time-series analysis tool in astronomy \citep{2009ApJ...698..895K,2014ApJ...788...33K,2018MNRAS.474.2094A,2020ApJ...895..122C,2021ApJ...907..105Y,2022ApJ...936...17H,2022MNRAS.513.2841C,2023ApJ...946...52Z,2023ARA&A..61..329A}. A GP can basically restore the true PSD from an astrophysical process without red noise leakage and aliasing. The PSD of the damped random walk (DRW) model is a natural BPL that is widely used to model the LCs of quasars \citep{2009ApJ...698..895K,2010ApJ...721.1014M,2013ApJ...765..106Z,2019PASP..131f3001M}. \citet{2017AJ....154..220F} have developed a tool---\texttt{celerite}\footnote{\url{https://celerite.readthedocs.io/en/stable/}}, that can implement GP modeling efficiently.

In this Letter, we use \texttt{celerite} to implement the DRW framework to model the long-term $\gamma$-ray LCs of blazars and the X-ray LCs of microquasars, using data from the Fermi-Large Area Telescope (LAT) and XMM-Newton X-ray telescope, respectively. Then, we show a scaling relationship between the characteristic timescales and the black hole masses. The remainder of this letter is structured as follows. In Section \ref{sec:data and method}, we describe the details of the sample data and the GPs. In Section \ref{sec:results and discussion}, we present the results of analysis of our sample, and discuss the scaling relationship involving the physics of the accretion disk.

\begin{deluxetable*}{cccccccccccc}[t] \label{tab:gamma}
	\tablenum{1}
	\tablecaption{Detailed Blazar Sample Information and Modeling Results}
	\setlength{\tabcolsep}{2pt}
	\tabletypesize{\scriptsize}
	\tablehead{
		\colhead{Fermi 4FGL Name} & \colhead{Identifier} & \colhead{Type} & \colhead{Redshift} & \colhead{$\delta_{D}$} & \colhead{log ($M_{BH}/M_{\sun}$)} & \colhead{Mean Cadence} & \colhead{Time Length} &\colhead{ln $\sigma_{DRW}$} & \colhead{ln $\tau_{DRW}$} & \colhead{ln $\tau_{rest}$} & Ref of $\delta_{D}$ and $M_{BH}$  \\ \colhead{} & \colhead{} & \colhead{} & \colhead{} & \colhead{} & \colhead{} & \colhead{(day)} & \colhead{(day)} & \colhead{} & \colhead{(day)} & \colhead{(day)} &  \\
	}
	\startdata
	4FGL J0221.1+3556  &  B2 0218+357  & FSRQ  & 0.944  &  2.24  &  $8.75\pm{0.24}$  & 23.48  & 5400        & $ 0.26^{+0.12}_{-0.10}$  & $4.82^{+0.28}_{-0.23}$ & $4.96^{+0.28}_{-0.23}$ & 1,2 \\
	4FGL J0222.6+4302* &  3C 66A       & BLL   & 0.37   &  6.23  &  $8.57\pm{0.6 }$  & 16.09  & $\sim$4650  & $-1.02^{+0.12}_{-0.10}$  & $4.48^{+0.28}_{-0.24}$ & $5.99^{+0.28}_{-0.24}$ & 3,4 \\
	4FGL J0238.6+1637  &  PKS 0235+164 & BLL   & 0.94   &  9.85  &  $8.58\pm{0.34}$  & 28.7   & 5310        & $ 0.14^{+0.15}_{-0.12}$  & $4.76^{+0.36}_{-0.29}$ & $6.38^{+0.36}_{-0.29}$ & 1,4 \\
	4FGL J0319.8+4130* &  NGC 1275     & BCU   & 0.018  &  1.33  &  $7.2 \pm{0.5 }$  & 15.2   & $\sim$4650  & $ 0.27^{+0.10}_{-0.08}$  & $4.27^{+0.23}_{-0.19}$ & $4.53^{+0.23}_{-0.19}$ & 3,2 \\
	4FGL J0334.2-4008  &  PKS 0332-403 & BLL   & 1.357  &  12.94 &  $8.67 \pm{0.82}$ & 29.43  & 5385        & $ 0.48^{+0.11}_{-0.10}$  & $4.21^{+0.38}_{-0.34}$ & $5.91^{+0.38}_{-0.34}$ & 1,4 \\
	4FGL J0428.6-3756* &  PKS 0426-380 & BLL   & 1.105  &  7.51  &  $8.77\pm{0.37}$  & 15.76  & $\sim$4650  & $-0.03^{+0.12}_{-0.10}$  & $4.65^{+0.27}_{-0.22}$ & $5.92^{+0.27}_{-0.22}$ & 1,4 \\
	4FGL J0509.4+0542* &  TXS 0506+056 & BLL   & 0.337  &  5.55  &  $8.5 \pm{0.6 }$  & 23.26  & $\sim$4650  & $-0.99^{+0.14}_{-0.11}$  & $4.70^{+0.37}_{-0.31}$ & $6.12^{+0.37}_{-0.31}$ & 3,4 \\
	4FGL J0538.8-4405* &  PKS 0537-441 & BLL   & 0.894  &  9.67  &  $8.45\pm{0.60}$  & 16.43  & $\sim$4650  & $ 0.09^{+0.21}_{-0.41}$  & $5.41^{+0.44}_{-0.31}$ & $7.04^{+0.44}_{-0.31}$ & 1,4 \\
	4FGL J0721.9+7120* &  S5 0716+71   & BLL   & 0.31   &  4.31  &   8.7             & 16.37  & $\sim$4650  & $-0.32^{+0.06}_{-0.05}$  & $3.18^{+0.16}_{-0.15}$ & $4.37^{+0.16}_{-0.15}$ & 3,4 \\
	4FGL J0809.8+5218  &  1ES 0806+524 & BLL   & 0.137  &  3.72  &  $8.55\pm{0.12}$  & 29.75  & 5295        & $-0.11^{+0.20}_{-0.15}$  & $5.47^{+0.52}_{-0.39}$ & $6.67^{+0.52}_{-0.39}$ & 1,4 \\
	4FGL J0811.4+0146  &  OJ 014       & BLL   & 1.148  &  21.79 &  $8.71\pm{0.69}$  & 39.2 8 & 4635        & $-2.21^{+0.14}_{-0.14}$  & $3.73^{+0.55}_{-0.55}$ & $6.04^{+0.55}_{-0.55}$ & 1,2 \\
	4FGL J0854.8+2006* &  OJ 287       & BLL   & 0.306  &  4.22  &  $8.8 \pm{0.5 }$  & 26.34  & $\sim$4650  & $-0.79^{+0.09}_{-0.08}$  & $3.73^{+0.23}_{-0.21}$ & $4.90^{+0.23}_{-0.21}$ & 3,4 \\
	4FGL J0957.6+5523  &  4C +55.17    & FSRQ  & 0.899  &  9.01  &  $8.8 \pm{0.06 }$ & 15.3   & 5400        & $-0.57^{+0.28}_{-0.18}$  & $5.76^{+0.95}_{-0.68}$ & $7.32^{+0.95}_{-0.68}$ & 1,4 \\
	4FGL J1058.4+0133  &  4C +01.28    & BLL   & 0.888  &  34.9  &  $9.5 \pm{0.09}$  & 25.79  & 5235        & $-1.11^{+0.10}_{-0.09}$  & $4.18^{+0.28}_{-0.25}$ & $7.09^{+0.28}_{-0.25}$ & 1,2 \\
	4FGL J1104.4+3812  &  Mrk 421      & BLL   & 0.03   &  3.32  &  $8.3 \pm{0.2}$   & 15.13  & 5400        & $-0.64^{+0.09}_{-0.08}$  & $4.21^{+0.20}_{-0.18}$ & $5.38^{+0.20}_{-0.18}$ & 3,4 \\
	4FGL J1127.0-1857  &  PKS 1124-186 & FSRQ  & 1.048  &  10.10 &  $8.9 \pm{0.22}$  & 27.4   & 3375        & $-0.30^{+0.21}_{-0.15}$  & $4.89^{+0.48}_{-0.36}$ & $6.48^{+0.48}_{-0.36}$ & 1,4 \\
	4FGL J1146.9+3958  &  S4 1144+40   & FSRQ  & 1.089  &  29.39 &  $9.04\pm{0.04}$  & 25.6   & 5235        & $-0.66^{+0.12}_{-0.10}$  & $4.45^{+0.29}_{-0.25}$ & $7.09^{+0.29}_{-0.25}$ & 1,2 \\
	4FGL J1159.5+2914* &  Ton 599      & FSRQ  & 0.725  &  6.76  &  $8.5 \pm{0.5 }$  & 22.04  & $\sim$4650  & $ 0.42^{+0.12}_{-0.10}$  & $4.30^{+0.26}_{-0.22}$ & $5.66^{+0.26}_{-0.22}$ & 3,4 \\
	4FGL J1217.9+3007  &  B2 1215+30   & BLL   & 0.237  &  3.04  &   8.39            & 17.2   & 5340        & $-1.62^{+0.12}_{-0.10}$  & $4.70^{+0.35}_{-0.29}$ & $5.59^{+0.35}_{-0.29}$ & 5,2 \\
	4FGL J1224.9+2122* &  4C +21.35    & FSRQ  & 0.433  &  5.00  &  $8.73\pm{0.02}$  & 24.13  & $\sim$4650  & $ 1.35^{+0.10}_{-0.09}$  & $4.02^{+0.23}_{-0.20}$ & $5.26^{+0.23}_{-0.20}$ & 1,4 \\
	4FGL J1229.0+0202* &  3C 273       & FSRQ  & 0.158  &  4.51  &  $8.9 \pm{0.5 }$  & 22.35  & $\sim$4650  & $ 1.00^{+0.07}_{-0.06}$  & $3.44^{+0.18}_{-0.17}$ & $4.79^{+0.18}_{-0.17}$ & 3,4 \\
	4FGL J1256.1-0547* &  3C 279       & FSRQ  & 0.536  &  7.18  &  $8.5 \pm{0.5 }$  & 15.71  & $\sim$4650  & $ 1.85^{+0.05}_{-0.05}$  & $3.06^{+0.15}_{-0.14}$ & $4.60^{+0.15}_{-0.14}$ & 3,4 \\
	4FGL J1504.4+1029* &  PKS 1502+106 & FSRQ  & 1.838  &  13.27 &  $9.13\pm{0.06}$  & 19.19  & $\sim$4650  & $ 0.91^{+0.24}_{-0.16}$  & $5.33^{+0.50}_{-0.34}$ & $6.87^{+0.50}_{-0.34}$ & 1,4 \\
	4FGL J1512.8-0906* &  PKS 1510-089 & FSRQ  & 0.36   &  4.89 &  $8.32\pm{0.13}$  & 15.66  & $\sim$4650  & $ 1.65^{+0.07}_{-0.06}$  & $3.68^{+0.17}_{-0.16}$ & $4.96^{+0.17}_{-0.16}$ & 1,4 \\	
	4FGL J1517.7-2422 &  AP Librae    & BLL   & 0.048  &  1.87  &  $9.09\pm{0.15}$  & 20.96  & 5325        & $ 1.16^{+0.10}_{-0.09}$  & $4.39^{+0.27}_{-0.24}$ & $4.97^{+0.27}_{-0.24}$ & 1,4 \\	
	4FGL J1522.1+3144  &  B2 1520+31   & FSRQ  & 1.488  &  35.22 &   9.4             & 17.5   & 4305        & $ 0.17^{+0.11}_{-0.09}$  & $4.36^{+0.26}_{-0.22}$ & $7.01^{+0.26}_{-0.22}$ & 3,2 \\
	4FGL J1555.7+1111* &  PG 1553+113  & BLL   & 0.36   &  7.49  &   8.7             & 15.15  & $\sim$4650  & $-2.12^{+0.27}_{-0.17}$  & $5.51^{+0.65}_{-0.43}$ & $7.21^{+0.65}_{-0.43}$ & 3,4 \\
	4FGL J1635.2+3808* &  4C +38.41    & FSRQ  & 1.814  &  20.04 &  $9.5 \pm{0.5 }$  & 17.22  & $\sim$4650  & $ 0.63^{+0.12}_{-0.10}$  & $4.60^{+0.27}_{-0.22}$ & $6.56^{+0.27}_{-0.22}$ & 3,4 \\
	4FGL J1653.8+3945  &  Mrk 501      & BLL   & 0.033  &  2.54  &   9.21            & 15.56  & 5400        & $-2.02^{+0.12}_{-0.10}$  & $4.65^{+0.30}_{-0.26}$ & $5.54^{+0.30}_{-0.26}$ & 6,4 \\
	4FGL J1806.8+6949  &  3C 371       & BLL   & 0.051  &  1.76  &  $7.1 \pm{0.31}$  & 27.17  & 5325        & $-2.25^{+0.12}_{-0.12}$  & $4.05^{+0.41}_{-0.39}$ & $4.56^{+0.41}_{-0.39}$ & 1,4 \\
	4FGL J2000.0+6508  &  1ES 1959+650 & BLL   & 0.018  &  2.81  &   8.09            & 17.31  & 5400        & $-1.71^{+0.21}_{-0.15}$  & $3.38^{+0.22}_{-0.20}$ & $4.39^{+0.22}_{-0.20}$ & 7,4 \\
	4FGL J2158.8-3013* &  PKS 2155-304 & BLL   & 0.117  &  5.03  &  $8.91\pm{0.22}$  & 15.35  & $\sim$4650  & $-1.07^{+0.10}_{-0.08}$  & $4.12^{+0.24}_{-0.20}$ & $5.62^{+0.24}_{-0.20}$ & 1,4 \\
	4FGL J2202.7+4216* &  BL Lacertae  & BLL   & 0.069  &  2.61  &  $8.5 \pm{0.2 }$  & 16.09  & $\sim$4650  & $ 0.91^{+0.10}_{-0.09}$  & $4.23^{+0.24}_{-0.20}$ & $5.12^{+0.24}_{-0.20}$ & 3,4 \\
	4FGL J2253.9+1609* &  3C 454.3     & FSRQ  & 0.859  &  12.51 &  $9.1 \pm{0.5 }$  & 16.37  & $\sim$4650  & $ 2.87^{+0.10}_{-0.09}$  & $4.30^{+0.22}_{-0.19}$ & $6.20^{+0.22}_{-0.19}$ & 3,4 \\
	\enddata
	\tablecomments{The superscript "*" indicates the modeling results from \citet{2022ApJ...930..157Z}, while the $\tau_{rest}$ of those sources are calculated by the Doppler factor corresponding to each source. FSRQ: flat-spectrum radio quasar; BLL: BL Lac object; BCU: blazar candidates of unknown. \\ References: (1) \citet{2021ApJS..253...46P}, (2) \citet{2018ApJ...866..137L}, (3) \citet{2022ApJ...930..157Z}, (4) \citet{2014RAA....14.1135F}, (5) \citet{2012NewA...17....8G}, (6) \citet{2012ApJ...759..114C}, (7) \citet{2012ApJ...752..157Z}.}
\end{deluxetable*}
\begin{deluxetable*}{ccccccccc}[t]
	\tablenum{2}
	\tablecaption{Detailed Microquasar Sample Information and Modeling Results. \label{tab:x}}
	\tablewidth{0pt}
	\setlength{\tabcolsep}{3pt}
	\tablehead{
		\colhead{Identifier} & \colhead{Obs ID} & \colhead{Observation Mode} & \colhead{log ($M_{BH}/M_{\sun}$)} & Mean Cadence (s) & Time Length (s) &\colhead{ln $\sigma_{DRW}$} & \colhead{ln $\tau_{DRW}$ (day)}
	}
	\startdata
	SS 433  & 0694870201 & Timing & $0.62 \pm{0.04}$ & 200 & 130800 & $-0.24_{-0.06}^{+0.06}$ & $-4.91^{+0.24}_{-0.22}$\\ \hline	
	GRO J1655-40  & 0112921301 & Burst & $0.73 \pm{0.01}$ & 217 & 41800 & $1.48_{-0.14}^{+0.18}$ & $-4.14^{+0.46}_{-0.37}$\\ \hline
	IGR J17091-3624 & 0743960201 & Timing & $1.10 \pm{0.09}$ & 200 & 58800 & $0.46_{-0.05}^{+0.06}$ & $-5.91^{+0.17}_{-0.17}$\\ \hline
	\multirow{2}{*}{LS I +61 303} & 0505981101 & Image & \multirow{2}{*}{$0.40 \pm{0.20}$} & 226 & 12200 & $-1.96^{+0.32}_{-0.21}$ & $-4.58^{+0.88}_{-0.63}$ \\
	& 0505981401 & Image &  & 217 & 12200 & $-2.42^{+0.32}_{-0.24}$ & $-4.61^{+0.95}_{-0.78}$ \\ \hline
	\multirow{3}{*}{GRS 1915+105} & 0144090101 & Timing & \multirow{3}{*}{$1.11 \pm{0.07}$} & 200 & 19000 & $2.07^{+0.11}_{-0.09}$ & $-5.67^{+0.31}_{-0.28}$ \\
	& 0851181701 & Timing &  & 293 & 35200 & $1.12^{+0.33}_{-0.32}$ & $-3.33^{+0.70}_{-0.48}$ \\
	& 0864960101 & Timing &  & 201 & 54000 & $0.34^{+0.09}_{-0.07}$ & $-4.91^{+0.20}_{-0.18}$ \\ \hline
	\multirow{3}{*}{H1743-322} & 0783540201 & Timing & \multirow{3}{*}{$1.04 \pm{0.07}$} & 200 & 139200 & $0.38^{+0.05}_{-0.05}$ & $-4.89^{+0.17}_{-0.16}$ \\
	& 0783540301 & Timing & & 200 & 136200 & $1.39^{+0.12}_{-0.10}$ & $-3.40^{+0.26}_{-0.21}$ \\
	& 0783540401 & Timing & & 200 & 131000 & $1.17^{+0.19}_{-0.13}$ & $-2.75^{+0.39}_{-0.29}$ \\ \hline
	\multirow{8}{*}{Cygnus X-1}	& 0202401201* & Burst  & \multirow{8}{*}{$1.17 \pm{0.02}$} & 200 & 17400 & $6.78^{+0.30}_{-0.19}$ & $-4.12^{+0.63}_{-0.40}$ \\
	& 0500880201 & Burst  & & 200 & 58600 & $5.40^{+0.20}_{-0.14}$ & $-3.47^{+0.42}_{-0.30}$ \\
	& 0605610401 & Timing & & 200 & 31600 & $3.81^{+0.09}_{-0.08}$ & $-5.39^{+0.23}_{-0.20}$ \\
	& 0745250201 & Timing & & 200 & 117000 & $5.53^{+0.13}_{-0.10}$ & $-3.41^{+0.27}_{-0.22}$ \\
	& 0745250501 & Timing & & 200 & 137600 & $4.92^{+0.08}_{-0.07}$ & $-4.15^{+0.18}_{-0.16}$ \\
	& 0745250601 & Timing & & 200 & 118000 & $4.38^{+0.06}_{-0.06}$ & $-4.62^{+0.14}_{-0.13}$ \\
	& 0745250701 & Timing & & 200 & 107000 & $5.17^{+0.12}_{-0.10}$ & $-3.62^{+0.25}_{-0.20}$ \\
	\enddata
	\tablecomments{References for the black hole masses: SS 433: \citet{2020AA...640A..96P}; GRO J1655-40: \citet{2014MNRAS.437.2554M}; IGR J17091-3624: \citet{2015ApJ...807..108I}; LS I +61 303: \citet{2011AA...527A...9Z}; GRS 1915+105: \citet{2013MNRAS.430.1832H}; H1743-322: \citet{2016cosp...41E1324M};  and Cygnus X-1: \citet{2011ApJ...742...84O}. The superscript "*" means that the spectral state is intermediate between soft and hard states. This is usually considered to be the critical state in which the radio jet begins to quench \citep{2006ARAA..44...49R}, but Cygnus X-1 still has a prominent radio flux \citep{2006AA...447..245W}. We thus assume that this observation is valid.}
\end{deluxetable*}

\section{Sample and Method} \label{sec:data and method}
\subsection{Blazar Sample}
To determine the scaling relationship and deal with the Doppler beaming effect, we take the overlap of bright sources in the black hole mass sample of Fermi blazars from \citet{2021ApJS..253...46P} and the Doppler factor estimate samples from \citet{2014RAA....14.1135F} and \citet{2018ApJ...866..137L} as our blazar sample. Our sample has 18 overlapping sources with those considered in \citet{2022ApJ...930..157Z}. We use their results of long-term LC modeling in our analysis. For sources with no modeling results, we extracted the long-term LC for each from Fermi-LAT.

Fermi-LAT has offered a new perspective of the extragalactic $\gamma$-ray sky \citep{2009ApJ...697.1071A}. The data in our sample were retrieved from the Fermi 4FGL database (i.e., Fermi Pass 8 database\footnote{\url{https://fermi.gsfc.nasa.gov/ssc/data/analysis/documentation/Pass8_usage.html}}; \citealt{2020ApJS..247...33A}). We utilized Fermitools 2.0.8\footnote{\url{https://github.com/fermi-lat/Fermitools-conda/}}, which is the official software for data reduction, to extract LCs with 15-day binning from 2008 August 4 to 2023 May 18 (54682-60082 MJD). First, we selected photon event files with an energy range of 0.1--300 GeV for binned maximum likelihood fitting. The spectra of all sample sources were fitted with a LogParabola model. The region of interest (ROI) was set as $15^{\circ}$. The filter expressions and maximum zenith angle were set to (DATA\_QUAL\textgreater0) \& (LAT\_CONFIG=1) and $90^{\circ}$, respectively. The parameters of the instrumental response function were selected based on P8R3\_SOURCE\_V3. The Galactic diffuse emission model file used in the likelihood analysis were gll\_iem\_v07.fits and iso\_P8R3\_SOURCE\_V3\_v1.txt, respectively. Finally, the LC data were generated using an unbinned likelihood analysis using the model files after binned likelihood fitting. We selected points with Test Statistic values exceeding 25 for the LC modeling.

\subsection{Microquasar Sample}
Since the origin of X-rays from microquasars remains uncertain \citep{1997ApJ...482..448N,2022Sci...378..650K}, we are interested here only in those sources which have reports of X-ray coming from jets, or show a correlation between radio and X-ray emissions. Our sample consisted of 30 microquasars selected from \citet{2002MNRAS.331..765B,2004RMxAC..20..203C,2006AA...447..245W,2006ApJ...650L.123G,2009ApJ...697..592T,2012BAAA...55..539V,2010MNRAS.403...61D,2014MNRAS.438L..41G,2018Natur.562...82A,2020ApJ...891...31X,2021A&A...647A.173K,arash_bahramian_2022_7059313}. We used the standard data reduction threads\footnote{\url{https://www.cosmos.esa.int/web/xmm-newton/sas-threads}} to extract the LCs in the hard state, excluding sources with no observations from XMM-Newton or those for which a sufficient number of data points could not be obtained from observations.

The XMM-Newton telescope enables high timing resolution at 0.2--12 KeV. For data reduction, we used the official data reduction tool--SAS 19.0.1 of XMM-Newton and standard calibration files. First, we used the \texttt{epproc} script to get calibrated and concatenated event lists from the PN sensor \citep{2001A&A...365L..18S}. After that, we used the \texttt{evselect} and \texttt{tabgtigen} scripts to obtain clean event files with background flares removed. The background count-rate threshold was set to "RATE\textless=0.4". In the Image mode, the area of the source was identified as a circle. For the Timing and Burst modes with ultra-high time resolution, the source was a rectangle with a central bright bar. Before generating the LCs, the \texttt{evselect} and \texttt{epatplot} scripts were used to subtract the background and check for pile-up, respectively. Finally, we used the \texttt{epiclccorr} script to generate usable LC files, where the data were grouped in 200 s bin$^{-1}$.

\subsection{A GP Model}
The variability of blazars can be described as a DRW process\footnote{Often called the Ornstein-Uhlenbeck process.} \citep{2012ApJ...760...51R,2022ApJ...930..157Z,2023ApJ...944..103Z}. The PSD of this model exhibits a BPL, which closely resembles the PSDs of blazars. In \texttt{celerite}, the covariance function of DRW can be written as:
\begin{equation}
	k(t_{n},t_{m})= 2\sigma^{2}_{DRW}\ \mathrm{exp}(-t_{nm}/\tau_{DRW}),
\end{equation}
where $t_{nm}=|t_{n}-t_{m}|$ is the time lag between measurements $m$ and $n$, $\sigma_{DRW}$ is the amplitude term, and $\tau_{DRW}$ is the characteristic timescale, which is related to the timescale of attenuation of producing the variability in the LC. The PSD of this model \citep{2017AJ....154..220F,2023ApJ...944..103Z} can be expressed as:
\begin{equation}
S(\omega)=\sqrt{\frac{8}{\pi}}\sigma^{2}_{DRW}\tau_{DRW}\frac{1}{1+(\omega\tau_{DRW})^{2}},
\end{equation}
where the relation between $f_{bend}$ in the BPL and the characteristic timescale is $\tau_{DRW}=1/(2\pi f_{bend})$.

We used the Markov Chain Monte Carlo (MCMC) sampler \texttt{emcee}\footnote{\url{https://emcee.readthedocs.io/en/v2.2.1/}} \citep{2013PASP..125..306F} to estimate the fit of the parameters of the model and the LCs of each source of our samples. In \texttt{emcee}, we use 32 parallel chains, each of which samples 10,000 steps for burn-in and 20,000 steps for generating the parameter distributions. If the model captures true variability, the standardized residual should be described as white noise, i.e. a Gaussian with $\mu=0$ and $\sigma=1$ \citep{2014ApJ...788...33K}. Meanwhile, the autocorrelation function (ACF) of the standardized residuals and squared standardized residuals should fall within the 95\% white noise confidence interval. Finally, we also check the posterior distributions of the parameters to ensure that the parameters are well converged.

\begin{deluxetable*}{ccc}[t]
	\tablenum{3}
	\tablecaption{Statistical Information of the Characteristic Timescale of the blazars and non-jetted SMBH accretion systems \citep{2021Sci...373..789B}. \label{tab:statistical}}
	\tablewidth{0pt}
	\setlength{\tabcolsep}{10pt}
	\tablehead{
		&	\colhead{Blazars} & \colhead{Non-jetted SMBH accretion systems}
	}
	\startdata
	$\tau_{min}$ (day)              & 79.04   & 2.00   \\
	$\tau_{max}$ (day)              & 1510.20 & 398.11 \\
	$\tau_{median}$ (day)           & 327.93  & 158.49 \\
    $\tau_{mean}$ (day)             & 485.68  & 167.15 \\
	Mean of log ($M_{BH}/M_{\sun}$) & 8.67    & 8.03   \\
	\enddata
\end{deluxetable*}

\section{Results and Discussion} \label{sec:results and discussion}
The standardized residuals of each source conformed to a Gaussian distribution in the results of modeling ($\mu=0$, $\sigma=1$). Both the ACFs of the residuals and the squared residuals fall within the 95\% white noise confidence interval. Although the DRW model fitted most sources well, characteristic timescales that are too large or too small can lead to deviations from real physical scenarios \citep{2017A&A...597A.128K,2021Sci...373..789B}. Therefore, we adopted the criteria of reliability of the characteristic timescale from \citet{2022ApJ...930..157Z}: the maximum characteristic timescale should not exceed 1/10 of the length of the LC and the minimum characteristic timescale should not be less than the average cadence of the observations. Finally, we obtained 34 blazars and seven microquasars through reliable modeling and estimated the black hole masses from our samples, as shown in Table \ref{tab:gamma} and Table \ref{tab:x}. The characteristic timescale of variability obtained by fitting is in the observer frame. Therefore, the characteristic timescale in the rest frame can be calculated by taking into account the Doppler beaming effect and the redshift of each source as:
\begin{equation}
	\tau_{rest}=\frac{\tau_{obs}}{1+z}\delta_{D},
\end{equation}
where $z$ is the redshift of the source and $\delta_{D}$ is the Doppler factor. For our sample, we used the jet Doppler factors estimated by \citet{2014RAA....14.1135F} and \citet{2018ApJ...866..137L} by limiting the pair-production optical depth and measuring the variability brightness temperature, respectively. Owing to a lack of results on the Doppler factor, and because the origin of the region of radiation remained uncertain, we did not consider the beaming effect for these seven microquasars. In addition, some studies suggest that this effect may be weak in microquasars \citep{2017ApJ...851..144L,2018Natur.562...82A}. In Table \ref{tab:statistical}, the characteristic timescales of the blazars are generally slightly longer than the timescales found by \citet{2021Sci...373..789B} for non-jetted SMBH accretion systems. This may occur because the mean black hole masses ($M_{BH}$) of our samples of blazars were larger than those identified for non-jetted accretion systems. The characteristic timescales of our blazar were similar to those reported in previous work (e.g., \citealt{2019ApJ...885...12R,2022ApJ...930..157Z}). The characteristic timescale of the seven microquasars were also consistent with the optical data on non-jetted SBH systems (see Table \ref{tab:x}; \citealt{2015SciA....1E0686S}).
\begin{figure*}[h]
	\figurenum{1}
	\plotone{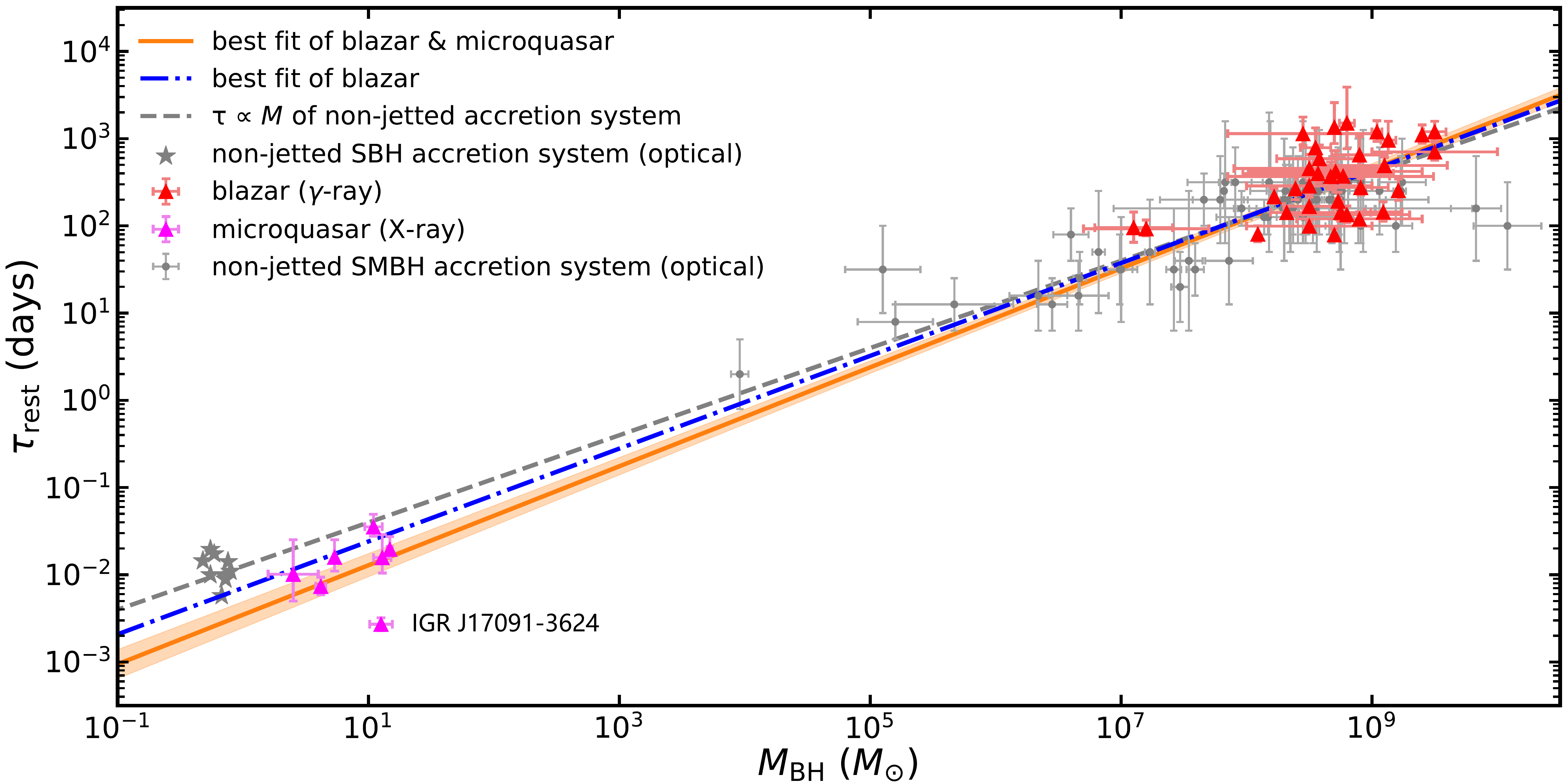}
	\caption{The mass-scaled damping timescale of our sample. The red and pink data points indicate the blazars and microquasars, respectively. The orange solid line and shaded region represent the best fit and 1 $\sigma$ confidence interval for all jetted systems, respectively. The blue dashed--dotted line shows a linear fit to the blazar data. The gray data points, star points, and dashed line represent the data and the fitted results for non-jetted systems from \citet{2021Sci...373..789B}. \label{fig:line}}
\end{figure*}

In Figure \ref{fig:line}, we obtained the $\tau \propto M_{BH}$ relation for the jetted systems using the \texttt{Linmix}\footnote{\url{https://github.com/jmeyers314/linmix}} software. The orange solid line and shaded region are the best linear fit for the jetted-systems and 1 $\sigma$ confidence interval for these points, respectively. The best-fitting result is:
\begin{equation}
	\tau_{rest}=120.47^{+14.84}_{-17.62} \ \text{days} \left( \frac{M_{BH}}{10^{8}M_{\sun}} \right)^{0.57^{+0.02}_{-0.02}},
\end{equation}

The additional $1\sigma$ intrinsic scatter of the data was $0.31\pm{0.05}$ and the Pearson correlation coefficient was 0.98. This linear relationship had a slope of 0.57, which is very similar to the slope ($\sim$0.5) of non-jetted systems observed at optical wavelengths (gray dashed line) in \citet{2021Sci...373..789B}. Even if we eliminate the microquasars from the fit, this linear relationship remains with a slope of 0.53, as shown by blue dashed--dotted line.
The data points of blazars in Figure \ref{fig:line} look somewhat scattered, possibly due to complex physical processes inside the jet that affect the shape of the LC, such as internal shock processes \citep{2010ApJ...711..445B}, local magnetic reconnection \citep{2009MNRAS.395L..29G},  jet-star interaction \citep{2012ApJ...749..119B}, and/or kink instabilities in the jet \citep{2017Galax...5...64N}.

To the 4 microquasars, LS I +61 303, GRS 1915+105, H1743-322 and Cygnus X-1, the multiple observations were shown in Table \ref{tab:x}. We calculated the $\tau \propto M_{BH}$ relation (equation (4)) with average and all timescales. The linear relationship of all timescales is almost same as the average values. Therefore, we only present the average results in the the Figure \ref{fig:line}.

Microquasar IGR J17091-3624 deviated from the fitting line (see Figure \ref{fig:line}), which means that the X-rays/radio correlation for this source might weak. In \citet{2011A&A...533L...4R}, the correlated radio/X-ray behaviour of microquasar IGR J17091-3624 was reported for the first time, only two radio points with the hard state X-ray luminosity consistent with the radio/X-ray correlation within the dispersion. The weak correlation suggests that the X-rays emission of IGR J17091-3624 does not come entirely from the jet.

\begin{figure}[h]
	\figurenum{2}
	\plotone{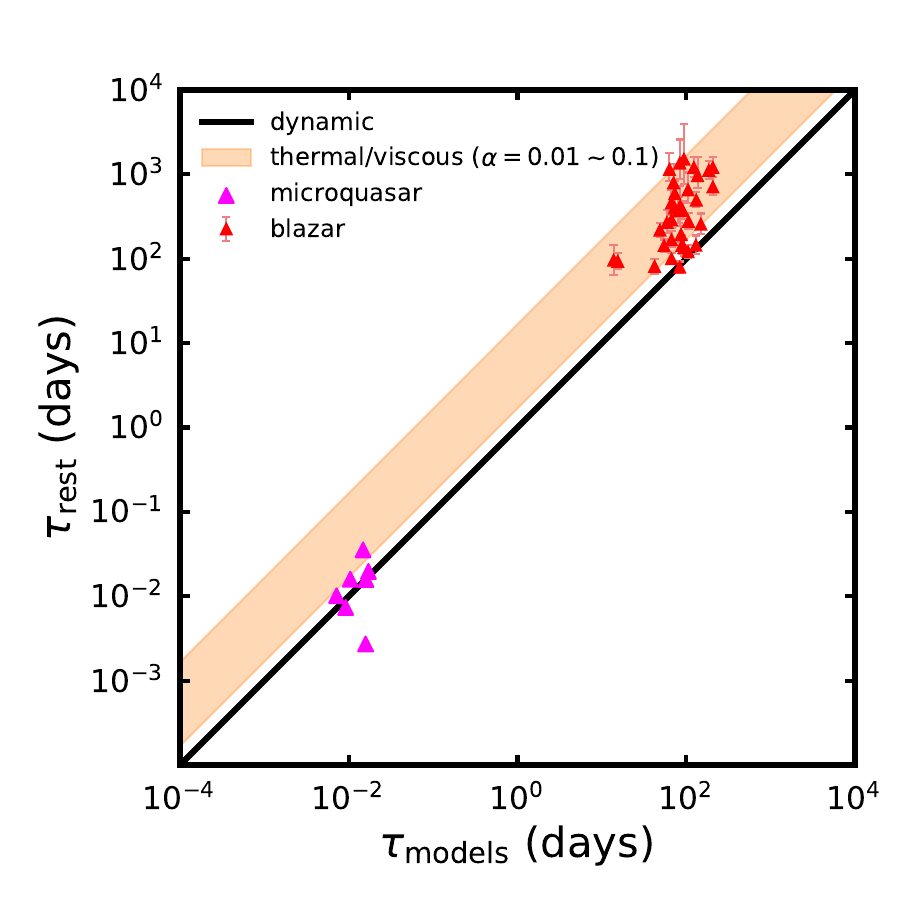}
	\caption{Characteristic timescales as a function of the model timescales. The black solid line and the orange shaded region indicate the dynamic and thermal timescales at the UV-emitting radius of the accretion disk, respectively. The upper and lower edges of the orange shaded region correspond to a viscosity parameter of 0.01 to 0.1, respectively. The red data points and pink triangles represent blazars and microquasars, respectively. \label{fig:models}}
\end{figure}

The scaling relationship shown in Figure \ref{fig:line} is very similar to that identified in non-jetted accretion systems by \citet{2021Sci...373..789B}. It can be explained by physical processes of the disk that influenced the variability of the jet. The characteristic annual timescale of the disk can be interpreted as thermal (related to restore thermal equilibrium), dynamic (related to orbital motion), or viscous (related to mass flow diffusion) timescales \citep{2006ASPC..360..265C,2021ApJ...907...96S,2021Sci...373..789B}. These timescales can be approximately written as:
\begin{equation}
	t_{th}=1680 \left( \frac{\alpha}{0.01} \right)^{-1} \times \left( \frac{M_{BH}}{10^{8}M_{\sun}} \right)  \left( \frac{R}{100R_{S}} \right)^{3/2} \text{days},
\end{equation}
and
\begin{equation}
	t_{th}\approx\alpha^{-1}t_{dyn}\approx(H/R)^{2}t_{vis}.
\end{equation}
where $\alpha$ is the viscosity parameter, $R_{S}$ is the Schwarzschild radius of the black hole, $R$ is the radial position, and $H$ is the disk thickness. In the standard disk model \citep{1973A&A....24..337S}, the viscous timescale is usually longer than the thermal timescale that we found here. In an ADAF \citep{1994ApJ...428L..13N}, given $H/R\sim$1, the viscous timescale and the thermal timescale are approximately equal \citep{2006ASPC..360..265C,2019MNRAS.483L..17D}. \citet{2021Sci...373..789B} suggested that the characteristic timescale of the optical emission originates from the thermal timescale or orbital timescale at the ultraviolet (UV)-emitting radii of the accretion disk, while our characteristic timescale are similar to theirs. Therefore, we assume that the characteristic timescale of the jet originates from variability at the UV-emitting radius of the accretion disk. \citet{2018ApJ...869..106M} used the microlensing variability technique to obtain the relationship between the UV-emitting radii of the accretion disk and the black hole mass, $\text{log}(R_{UV}/\text{cm})=(15.85\pm{0.12}) + (0.66\pm{0.15})\text{log}(M_{BH}/10^{9}M_{\sun})$. Simulations of magneto-rotational instability have shown that the viscosity parameter of standard disks and ADAFs rangs from 0.01 and 0.1 \citep{1999AcA....49..391S,2001ApJ...548..348H,2012MNRAS.426.1107L,2013ApJ...772..102H}. In Figure \ref{fig:models}, our $\gamma$-ray data conforms to the thermal/viscosity timescale model (orange shaded region) with viscosity parameters in range 0.01 $\sim$ 0.1. The timescales of seven microquasars prefer the dynamic model, but thermal model cannot be ruled out.

Microquasars and blazars exhibit a same relationship, proposed that the hard state X-ray emissions of seven microquasars are related with their jets. Further X-ray observations of microquasars could reveal the origins of X-rays and the underlying physical processes.

In the theoretical framework, the magnetic field is enhanced near the inner disk by the dragging of accretion material, which is in balance with gravity, forming a magnetically arrested disk in an ADAF \citep{2003ApJ...592.1042I,2003PASJ...55L..69N,2008ApJ...677..317I}. This scenario is thought to be the generator of radio jet flares \citep{2014ARA&A..52..529Y,2022A&A...664A.166R,2023Sci...381..961Y}.
The relation between the UV-emitting radii and the inner region near the black hole is a complex problem. In SMBH systems, the UV-emitting radii and corona (related to jet) would be coupled with the magnetic field. The variability can then propagate from UV-emitting radii to corona via Alfv\'{e}n waves, which can travel close to the light speed \citep{2020ApJ...891..178S}. In our work, both the microquasars and blazars follow the same mass scaling relationship, have the similar jetted system. It is naturally to infer that the mechanism of their characteristic timescales could be similar. When this variabilitiy propagates to the inner ADAF region of the accretion disk, they may exert an influence on the jet. The specific mechanism of transmission of this influence to the jet remains unclear. In addition, physical processes occurring only inside the jet may also produce characteristic timescales. However, \citet{2014ApJ...791...21F} utilized frequency-domain leptonic simulations, revealing that the characteristic timescales within blazar jets are primarily of the order of several hours to a few days, which is considerably shorter than the timescales calculated here.

In conclusion, we discovered the mass-scaled damping timescale relationship, ranging from microquasars to blazars. The results help us better understand the physical mechanism of accretion disks and jets. They also indirectly support the view that the X-ray emissions of microquasars are related to the jet. Furthermore, the variability at UV-emitting radii of the accretion disk may be transmitted to the base of the jet, and thus contribute to its characteristic timescale. This suggests a possible disk--jet connection in jetted black hole systems. We expect future multi-band observations of the microquasars, especially X-ray emissions, to unravel this mystery.

\begin{acknowledgments}
\centerline{Acknowledgments}
We thank the reviewer for providing valuable comments that significantly improved this work. This work was partly supported by the National Science Foundation of China (grant Nos. 12263007 and 12233006), the High-level talent support program of Yunnan Province. This work is based on observations conducted by XMM-Newton, an ESA science mission with instruments and contributions directly funded by ESA Member States and the USA (NASA).
\end{acknowledgments}

\facilities{Fermi (LAT) and XMM-Newton (EPIC).}
\software{\texttt{celerite} \citep{2017AJ....154..220F}, \texttt{corner} \citep{corner}, \texttt{emcee} \citep{2013PASP..125..306F}, \texttt{linmix} \citep{2007ApJ...665.1489K}, \texttt{Numpy} \citep{2020Natur.585..357H} and \texttt{Matplotlib} \citep{2007CSE.....9...90H}.}

\bibliography{d-j}{}

\begin{thebibliography}{}
\expandafter\ifx\csname natexlab\endcsname\relax\def\natexlab#1{#1}\fi
\providecommand{\url}[1]{\href{#1}{#1}}
\providecommand{\dodoi}[1]{doi:~\href{http://doi.org/#1}{\nolinkurl{#1}}}
\providecommand{\doeprint}[1]{\href{http://ascl.net/#1}{\nolinkurl{http://ascl.net/#1}}}
\providecommand{\doarXiv}[1]{\href{https://arxiv.org/abs/#1}{\nolinkurl{https://arxiv.org/abs/#1}}}

\bibitem[{{Abdollahi} {et~al.}(2020){Abdollahi}, {Acero}, {Ackermann},
  {Ajello}, {Atwood}, {Axelsson}, {Baldini}, {Ballet}, {Barbiellini},
  {Bastieri}, {Becerra Gonzalez}, {Bellazzini}, {Berretta}, {Bissaldi},
  {Blandford}, {Bloom}, {Bonino}, {Bottacini}, {Brandt}, {Bregeon}, {Bruel},
  {Buehler}, {Burnett}, {Buson}, {Cameron}, {Caputo}, {Caraveo}, {Casandjian},
  {Castro}, {Cavazzuti}, {Charles}, {Chaty}, {Chen}, {Cheung}, {Chiaro},
  {Ciprini}, {Cohen-Tanugi}, {Cominsky}, {Coronado-Bl{\'a}zquez}, {Costantin},
  {Cuoco}, {Cutini}, {D'Ammando}, {DeKlotz}, {de la Torre Luque}, {de Palma},
  {Desai}, {Digel}, {Di Lalla}, {Di Mauro}, {Di Venere}, {Dom{\'\i}nguez},
  {Dumora}, {Fana Dirirsa}, {Fegan}, {Ferrara}, {Franckowiak}, {Fukazawa},
  {Funk}, {Fusco}, {Gargano}, {Gasparrini}, {Giglietto}, {Giommi}, {Giordano},
  {Giroletti}, {Glanzman}, {Green}, {Grenier}, {Griffin}, {Grondin}, {Grove},
  {Guiriec}, {Harding}, {Hayashi}, {Hays}, {Hewitt}, {Horan},
  {J{\'o}hannesson}, {Johnson}, {Kamae}, {Kerr}, {Kocevski}, {Kovac'evic'},
  {Kuss}, {Landriu}, {Larsson}, {Latronico}, {Lemoine-Goumard}, {Li},
  {Liodakis}, {Longo}, {Loparco}, {Lott}, {Lovellette}, {Lubrano}, {Madejski},
  {Maldera}, {Malyshev}, {Manfreda}, {Marchesini}, {Marcotulli},
  {Mart{\'\i}-Devesa}, {Martin}, {Massaro}, {Mazziotta}, {McEnery}, {Mereu},
  {Meyer}, {Michelson}, {Mirabal}, {Mizuno}, {Monzani}, {Morselli},
  {Moskalenko}, {Negro}, {Nuss}, {Ojha}, {Omodei}, {Orienti}, {Orlando},
  {Ormes}, {Palatiello}, {Paliya}, {Paneque}, {Pei}, {Pe{\~n}a-Herazo},
  {Perkins}, {Persic}, {Pesce-Rollins}, {Petrosian}, {Petrov}, {Piron}, {Poon},
  {Porter}, {Principe}, {Rain{\`o}}, {Rando}, {Razzano}, {Razzaque}, {Reimer},
  {Reimer}, {Remy}, {Reposeur}, {Romani}, {Saz Parkinson}, {Schinzel},
  {Serini}, {Sgr{\`o}}, {Siskind}, {Smith}, {Spandre}, {Spinelli}, {Strong},
  {Suson}, {Tajima}, {Takahashi}, {Tak}, {Thayer}, {Thompson}, {Tibaldo},
  {Torres}, {Torresi}, {Valverde}, {Van Klaveren}, {van Zyl}, {Wood},
  {Yassine}, \& {Zaharijas}}]{2020ApJS..247...33A}
{Abdollahi}, S., {Acero}, F., {Ackermann}, M., {et~al.} 2020, \apjs, 247, 33,
  \dodoi{10.3847/1538-4365/ab6bcb}

\bibitem[{{Abeysekara} {et~al.}(2018){Abeysekara}, {Albert}, {Alfaro},
  {Alvarez}, {{\'A}lvarez}, {Arceo}, {Arteaga-Vel{\'a}zquez}, {Avila Rojas},
  {Ayala Solares}, {Belmont-Moreno}, {BenZvi}, {Brisbois}, {Caballero-Mora},
  {Capistr{\'a}n}, {Carrami{\~n}ana}, {Casanova}, {Castillo}, {Cotti},
  {Cotzomi}, {Couti{\~n}o de Le{\'o}n}, {De Le{\'o}n}, {De la Fuente},
  {D{\'\i}az-V{\'e}lez}, {Dichiara}, {Dingus}, {DuVernois}, {Ellsworth},
  {Engel}, {Espinoza}, {Fang}, {Fleischhack}, {Fraija}, {Galv{\'a}n-G{\'a}mez},
  {Garc{\'\i}a-Gonz{\'a}lez}, {Garfias}, {Gonz{\'a}lez-Mu{\~n}oz},
  {Gonz{\'a}lez}, {Goodman}, {Hampel-Arias}, {Harding}, {Hernandez}, {Hinton},
  {Hona}, {Hueyotl-Zahuantitla}, {Hui}, {H{\"u}ntemeyer}, {Iriarte},
  {Jardin-Blicq}, {Joshi}, {Kaufmann}, {Kar}, {Kunde}, {Lauer}, {Lee},
  {Le{\'o}n Vargas}, {Li}, {Linnemann}, {Longinotti}, {Luis-Raya},
  {L{\'o}pez-Coto}, {Malone}, {Marinelli}, {Martinez}, {Martinez-Castellanos},
  {Mart{\'\i}nez-Castro}, {Matthews}, {Miranda-Romagnoli}, {Moreno},
  {Mostaf{\'a}}, {Nayerhoda}, {Nellen}, {Newbold}, {Nisa}, {Noriega-Papaqui},
  {Pretz}, {P{\'e}rez-P{\'e}rez}, {Ren}, {Rho}, {Rivi{\`e}re},
  {Rosa-Gonz{\'a}lez}, {Rosenberg}, {Ruiz-Velasco}, {Salesa Greus}, {Sandoval},
  {Schneider}, {Schoorlemmer}, {Seglar Arroyo}, {Sinnis}, {Smith}, {Springer},
  {Surajbali}, {Taboada}, {Tibolla}, {Tollefson}, {Torres}, {Vianello},
  {Villase{\~n}or}, {Weisgarber}, {Werner}, {Westerhoff}, {Wood}, {Yapici},
  {Yodh}, {Zepeda}, {Zhang}, \& {Zhou}}]{2018Natur.562...82A}
{Abeysekara}, A.~U., {Albert}, A., {Alfaro}, R., {et~al.} 2018, \nat, 562, 82,
  \dodoi{10.1038/s41586-018-0565-5}

\bibitem[{{Aigrain} \& {Foreman-Mackey}(2023)}]{2023ARA&A..61..329A}
{Aigrain}, S., \& {Foreman-Mackey}, D. 2023, \araa, 61, 329,
  \dodoi{10.1146/annurev-astro-052920-103508}

\bibitem[{{Angus} {et~al.}(2018){Angus}, {Morton}, {Aigrain}, {Foreman-Mackey},
  \& {Rajpaul}}]{2018MNRAS.474.2094A}
{Angus}, R., {Morton}, T., {Aigrain}, S., {Foreman-Mackey}, D., \& {Rajpaul},
  V. 2018, \mnras, 474, 2094, \dodoi{10.1093/mnras/stx2109}

\bibitem[{{Atwood} {et~al.}(2009){Atwood}, {Abdo}, {Ackermann}, {Althouse},
  {Anderson}, {Axelsson}, {Baldini}, {Ballet}, {Band}, {Barbiellini},
  {Bartelt}, {Bastieri}, {Baughman}, {Bechtol}, {B{\'e}d{\'e}r{\`e}de},
  {Bellardi}, {Bellazzini}, {Berenji}, {Bignami}, {Bisello}, {Bissaldi},
  {Blandford}, {Bloom}, {Bogart}, {Bonamente}, {Bonnell}, {Borgland},
  {Bouvier}, {Bregeon}, {Brez}, {Brigida}, {Bruel}, {Burnett}, {Busetto},
  {Caliandro}, {Cameron}, {Caraveo}, {Carius}, {Carlson}, {Casandjian},
  {Cavazzuti}, {Ceccanti}, {Cecchi}, {Charles}, {Chekhtman}, {Cheung},
  {Chiang}, {Chipaux}, {Cillis}, {Ciprini}, {Claus}, {Cohen-Tanugi},
  {Condamoor}, {Conrad}, {Corbet}, {Corucci}, {Costamante}, {Cutini}, {Davis},
  {Decotigny}, {DeKlotz}, {Dermer}, {de Angelis}, {Digel}, {do Couto e Silva},
  {Drell}, {Dubois}, {Dumora}, {Edmonds}, {Fabiani}, {Farnier}, {Favuzzi},
  {Flath}, {Fleury}, {Focke}, {Funk}, {Fusco}, {Gargano}, {Gasparrini},
  {Gehrels}, {Gentit}, {Germani}, {Giebels}, {Giglietto}, {Giommi}, {Giordano},
  {Glanzman}, {Godfrey}, {Grenier}, {Grondin}, {Grove}, {Guillemot}, {Guiriec},
  {Haller}, {Harding}, {Hart}, {Hays}, {Healey}, {Hirayama}, {Hjalmarsdotter},
  {Horn}, {Hughes}, {J{\'o}hannesson}, {Johansson}, {Johnson}, {Johnson},
  {Johnson}, {Johnson}, {Kamae}, {Katagiri}, {Kataoka}, {Kavelaars}, {Kawai},
  {Kelly}, {Kerr}, {Klamra}, {Kn{\"o}dlseder}, {Kocian}, {Komin}, {Kuehn},
  {Kuss}, {Landriu}, {Latronico}, {Lee}, {Lee}, {Lemoine-Goumard}, {Lionetto},
  {Longo}, {Loparco}, {Lott}, {Lovellette}, {Lubrano}, {Madejski}, {Makeev},
  {Marangelli}, {Massai}, {Mazziotta}, {McEnery}, {Menon}, {Meurer},
  {Michelson}, {Minuti}, {Mirizzi}, {Mitthumsiri}, {Mizuno}, {Moiseev},
  {Monte}, {Monzani}, {Moretti}, {Morselli}, {Moskalenko}, {Murgia},
  {Nakamori}, {Nishino}, {Nolan}, {Norris}, {Nuss}, {Ohno}, {Ohsugi}, {Omodei},
  {Orlando}, {Ormes}, {Paccagnella}, {Paneque}, {Panetta}, {Parent}, {Pearce},
  {Pepe}, {Perazzo}, {Pesce-Rollins}, {Picozza}, {Pieri}, {Pinchera}, {Piron},
  {Porter}, {Poupard}, {Rain{\`o}}, {Rando}, {Rapposelli}, {Razzano}, {Reimer},
  {Reimer}, {Reposeur}, {Reyes}, {Ritz}, {Rochester}, {Rodriguez}, {Romani},
  {Roth}, {Russell}, {Ryde}, {Sabatini}, {Sadrozinski}, {Sanchez}, {Sander},
  {Sapozhnikov}, {Parkinson}, {Scargle}, {Schalk}, {Scolieri}, {Sgr{\`o}},
  {Share}, {Shaw}, {Shimokawabe}, {Shrader}, {Sierpowska-Bartosik}, {Siskind},
  {Smith}, {Smith}, {Spandre}, {Spinelli}, {Starck}, {Stephens}, {Strickman},
  {Strong}, {Suson}, {Tajima}, {Takahashi}, {Takahashi}, {Tanaka}, {Tenze},
  {Tether}, {Thayer}, {Thayer}, {Thompson}, {Tibaldo}, {Tibolla}, {Torres},
  {Tosti}, {Tramacere}, {Turri}, {Usher}, {Vilchez}, {Vitale}, {Wang},
  {Watters}, {Winer}, {Wood}, {Ylinen}, \& {Ziegler}}]{2009ApJ...697.1071A}
{Atwood}, W.~B., {Abdo}, A.~A., {Ackermann}, M., {et~al.} 2009, \apj, 697,
  1071, \dodoi{10.1088/0004-637X/697/2/1071}

\bibitem[{Bahramian \& Rushton(2022)}]{arash_bahramian_2022_7059313}
Bahramian, A., \& Rushton, A. 2022, bersavosh/XRB-LrLx\_pub: update 20220908,
  v220908,  Zenodo, \dodoi{10.5281/zenodo.7059313}

\bibitem[{{Barkov} {et~al.}(2012){Barkov}, {Aharonian}, {Bogovalov}, {Kelner},
  \& {Khangulyan}}]{2012ApJ...749..119B}
{Barkov}, M.~V., {Aharonian}, F.~A., {Bogovalov}, S.~V., {Kelner}, S.~R., \&
  {Khangulyan}, D. 2012, \apj, 749, 119, \dodoi{10.1088/0004-637X/749/2/119}

\bibitem[{{Bellan}(2018)}]{2018PPCF...60a4006B}
{Bellan}, P.~M. 2018, Plasma Physics and Controlled Fusion, 60, 014006,
  \dodoi{10.1088/1361-6587/aa85f9}

\bibitem[{{Belloni} {et~al.}(2000){Belloni}, {Klein-Wolt}, {M{\'e}ndez}, {van
  der Klis}, \& {van Paradijs}}]{2000A&A...355..271B}
{Belloni}, T., {Klein-Wolt}, M., {M{\'e}ndez}, M., {van der Klis}, M., \& {van
  Paradijs}, J. 2000, \aap, 355, 271, \dodoi{10.48550/arXiv.astro-ph/0001103}

\bibitem[{{Blandford} {et~al.}(2019){Blandford}, {Meier}, \&
  {Readhead}}]{2019ARA&A..57..467B}
{Blandford}, R., {Meier}, D., \& {Readhead}, A. 2019, \araa, 57, 467,
  \dodoi{10.1146/annurev-astro-081817-051948}

\bibitem[{{Blandford} \& {Payne}(1982)}]{1982MNRAS.199..883B}
{Blandford}, R.~D., \& {Payne}, D.~G. 1982, \mnras, 199, 883,
  \dodoi{10.1093/mnras/199.4.883}

\bibitem[{{Blandford} \& {Znajek}(1977)}]{1977MNRAS.179..433B}
{Blandford}, R.~D., \& {Znajek}, R.~L. 1977, \mnras, 179, 433,
  \dodoi{10.1093/mnras/179.3.433}

\bibitem[{{B{\"o}ttcher} \& {Dermer}(2010)}]{2010ApJ...711..445B}
{B{\"o}ttcher}, M., \& {Dermer}, C.~D. 2010, \apj, 711, 445,
  \dodoi{10.1088/0004-637X/711/1/445}

\bibitem[{{Brocksopp} {et~al.}(2002){Brocksopp}, {Fender}, {McCollough},
  {Pooley}, {Rupen}, {Hjellming}, {de la Force}, {Spencer}, {Muxlow},
  {Garrington}, \& {Trushkin}}]{2002MNRAS.331..765B}
{Brocksopp}, C., {Fender}, R.~P., {McCollough}, M., {et~al.} 2002, \mnras, 331,
  765, \dodoi{10.1046/j.1365-8711.2002.05230.x}

\bibitem[{{Burke} {et~al.}(2021){Burke}, {Shen}, {Blaes}, {Gammie}, {Horne},
  {Jiang}, {Liu}, {McHardy}, {Morgan}, {Scaringi}, \&
  {Yang}}]{2021Sci...373..789B}
{Burke}, C.~J., {Shen}, Y., {Blaes}, O., {et~al.} 2021, Science, 373, 789,
  \dodoi{10.1126/science.abg9933}

\bibitem[{{Cai} {et~al.}(2022){Cai}, {Kurtanidze}, {Liu}, {Kurtanidze},
  {Nikolashvili}, {Xiao}, \& {Fan}}]{2022ApJS..260...47C}
{Cai}, J.~T., {Kurtanidze}, S.~O., {Liu}, Y., {et~al.} 2022, \apjs, 260, 47,
  \dodoi{10.3847/1538-4365/ac666b}

\bibitem[{{Chai} {et~al.}(2012){Chai}, {Cao}, \& {Gu}}]{2012ApJ...759..114C}
{Chai}, B., {Cao}, X., \& {Gu}, M. 2012, \apj, 759, 114,
  \dodoi{10.1088/0004-637X/759/2/114}

\bibitem[{{Choudhury} \& {Rao}(2004)}]{2004RMxAC..20..203C}
{Choudhury}, M., \& {Rao}, A.~R. 2004, in Revista Mexicana de Astronomia y
  Astrofisica Conference Series, Vol.~20, Revista Mexicana de Astronomia y
  Astrofisica Conference Series, ed. G.~{Tovmassian} \& E.~{Sion}, 203--203,
  \dodoi{10.48550/arXiv.astro-ph/0312601}

\bibitem[{{Corbel} {et~al.}(2002){Corbel}, {Fender}, {Tzioumis}, {Tomsick},
  {Orosz}, {Miller}, {Wijnands}, \& {Kaaret}}]{2002Sci...298..196C}
{Corbel}, S., {Fender}, R.~P., {Tzioumis}, A.~K., {et~al.} 2002, Science, 298,
  196, \dodoi{10.1126/science.1075857}

\bibitem[{{Corbel} {et~al.}(2003){Corbel}, {Nowak}, {Fender}, {Tzioumis}, \&
  {Markoff}}]{2003A&A...400.1007C}
{Corbel}, S., {Nowak}, M.~A., {Fender}, R.~P., {Tzioumis}, A.~K., \& {Markoff},
  S. 2003, \aap, 400, 1007, \dodoi{10.1051/0004-6361:20030090}

\bibitem[{{Covino} {et~al.}(2020){Covino}, {Landoni}, {Sandrinelli}, \&
  {Treves}}]{2020ApJ...895..122C}
{Covino}, S., {Landoni}, M., {Sandrinelli}, A., \& {Treves}, A. 2020, \apj,
  895, 122, \dodoi{10.3847/1538-4357/ab8bd4}

\bibitem[{{Covino} {et~al.}(2022){Covino}, {Tobar}, \&
  {Treves}}]{2022MNRAS.513.2841C}
{Covino}, S., {Tobar}, F., \& {Treves}, A. 2022, \mnras, 513, 2841,
  \dodoi{10.1093/mnras/stac596}

\bibitem[{{Czerny}(2006)}]{2006ASPC..360..265C}
{Czerny}, B. 2006, in Astronomical Society of the Pacific Conference Series,
  Vol. 360, AGN Variability from X-Rays to Radio Waves, ed. C.~M. {Gaskell},
  I.~M. {McHardy}, B.~M. {Peterson}, \& S.~G. {Sergeev}, 265

\bibitem[{{Czerny} {et~al.}(1999){Czerny}, {Schwarzenberg-Czerny}, \&
  {Loska}}]{1999MNRAS.303..148C}
{Czerny}, B., {Schwarzenberg-Czerny}, A., \& {Loska}, Z. 1999, \mnras, 303,
  148, \dodoi{10.1046/j.1365-8711.1999.02196.x}

\bibitem[{{Dai} {et~al.}(2001){Dai}, {Xie}, {Li}, {Zhou}, {Liu}, \&
  {Jiang}}]{2001AJ....122.2901D}
{Dai}, B.~Z., {Xie}, G.~Z., {Li}, K.~H., {et~al.} 2001, \aj, 122, 2901,
  \dodoi{10.1086/324450}

\bibitem[{{Dai} {et~al.}(2009){Dai}, {Li}, {Liu}, {Zhang}, {Na}, {Wu}, {Hao},
  {Xiang}, {Jiang}, \& {Zhang}}]{2009MNRAS.392.1181D}
{Dai}, B.~Z., {Li}, X.~H., {Liu}, Z.~M., {et~al.} 2009, \mnras, 392, 1181,
  \dodoi{10.1111/j.1365-2966.2008.14137.x}

\bibitem[{{Dai} {et~al.}(2015){Dai}, {Zeng}, {Jiang}, {Fan}, {Hu}, {Zhang},
  {Yang}, {Yan}, {Wang}, \& {Zhang}}]{2015ApJS..218...18D}
{Dai}, B.-z., {Zeng}, W., {Jiang}, Z.-j., {et~al.} 2015, \apjs, 218, 18,
  \dodoi{10.1088/0067-0049/218/2/18}

\bibitem[{{Dexter} \& {Begelman}(2019)}]{2019MNRAS.483L..17D}
{Dexter}, J., \& {Begelman}, M.~C. 2019, \mnras, 483, L17,
  \dodoi{10.1093/mnrasl/sly213}

\bibitem[{{Done} {et~al.}(2007){Done}, {Gierli{\'n}ski}, \&
  {Kubota}}]{2007A&ARv..15....1D}
{Done}, C., {Gierli{\'n}ski}, M., \& {Kubota}, A. 2007, \aapr, 15, 1,
  \dodoi{10.1007/s00159-007-0006-1}

\bibitem[{{Dunn} {et~al.}(2010){Dunn}, {Fender}, {K{\"o}rding}, {Belloni}, \&
  {Cabanac}}]{2010MNRAS.403...61D}
{Dunn}, R.~J.~H., {Fender}, R.~P., {K{\"o}rding}, E.~G., {Belloni}, T., \&
  {Cabanac}, C. 2010, \mnras, 403, 61, \dodoi{10.1111/j.1365-2966.2010.16114.x}

\bibitem[{{Fan} {et~al.}(2014){Fan}, {Bastieri}, {Yang}, {Liu}, {Hua}, {Yuan},
  \& {Wu}}]{2014RAA....14.1135F}
{Fan}, J.-H., {Bastieri}, D., {Yang}, J.-H., {et~al.} 2014, Research in
  Astronomy and Astrophysics, 14, 1135, \dodoi{10.1088/1674-4527/14/9/004}

\bibitem[{{Fan} {et~al.}(2021){Fan}, {Kurtanidze}, {Liu}, {Kurtanidze},
  {Nikolashvili}, {Liu}, {Zhang}, {Cai}, {Zhu}, {He}, {Yang}, {Yang}, {Gu},
  {Luo}, \& {Yuan}}]{2021ApJS..253...10F}
{Fan}, J.~H., {Kurtanidze}, S.~O., {Liu}, Y., {et~al.} 2021, \apjs, 253, 10,
  \dodoi{10.3847/1538-4365/abd32d}

\bibitem[{{Finke} \& {Becker}(2014)}]{2014ApJ...791...21F}
{Finke}, J.~D., \& {Becker}, P.~A. 2014, \apj, 791, 21,
  \dodoi{10.1088/0004-637X/791/1/21}

\bibitem[{Foreman-Mackey(2016)}]{corner}
Foreman-Mackey, D. 2016, The Journal of Open Source Software, 1, 24,
  \dodoi{10.21105/joss.00024}

\bibitem[{{Foreman-Mackey} {et~al.}(2017){Foreman-Mackey}, {Agol},
  {Ambikasaran}, \& {Angus}}]{2017AJ....154..220F}
{Foreman-Mackey}, D., {Agol}, E., {Ambikasaran}, S., \& {Angus}, R. 2017, \aj,
  154, 220, \dodoi{10.3847/1538-3881/aa9332}

\bibitem[{{Foreman-Mackey} {et~al.}(2013){Foreman-Mackey}, {Hogg}, {Lang}, \&
  {Goodman}}]{2013PASP..125..306F}
{Foreman-Mackey}, D., {Hogg}, D.~W., {Lang}, D., \& {Goodman}, J. 2013, \pasp,
  125, 306, \dodoi{10.1086/670067}

\bibitem[{{Gallo} {et~al.}(2014){Gallo}, {Plotkin}, \&
  {Jonker}}]{2014MNRAS.438L..41G}
{Gallo}, E., {Plotkin}, R.~M., \& {Jonker}, P.~G. 2014, \mnras, 438, L41,
  \dodoi{10.1093/mnrasl/slt152}

\bibitem[{{Giannios} {et~al.}(2009){Giannios}, {Uzdensky}, \&
  {Begelman}}]{2009MNRAS.395L..29G}
{Giannios}, D., {Uzdensky}, D.~A., \& {Begelman}, M.~C. 2009, \mnras, 395, L29,
  \dodoi{10.1111/j.1745-3933.2009.00635.x}

\bibitem[{{Gupta} \& {B{\"o}ttcher}(2006)}]{2006ApJ...650L.123G}
{Gupta}, S., \& {B{\"o}ttcher}, M. 2006, \apjl, 650, L123,
  \dodoi{10.1086/508880}

\bibitem[{{Gupta} {et~al.}(2012){Gupta}, {Pandey}, {Singh}, {Rani}, {Pan},
  {Fan}, \& {Gupta}}]{2012NewA...17....8G}
{Gupta}, S.~P., {Pandey}, U.~S., {Singh}, K., {et~al.} 2012, \na, 17, 8,
  \dodoi{10.1016/j.newast.2011.05.005}

\bibitem[{{Harris} {et~al.}(2020){Harris}, {Millman}, {van der Walt},
  {Gommers}, {Virtanen}, {Cournapeau}, {Wieser}, {Taylor}, {Berg}, {Smith},
  {Kern}, {Picus}, {Hoyer}, {van Kerkwijk}, {Brett}, {Haldane}, {del R{\'\i}o},
  {Wiebe}, {Peterson}, {G{\'e}rard-Marchant}, {Sheppard}, {Reddy}, {Weckesser},
  {Abbasi}, {Gohlke}, \& {Oliphant}}]{2020Natur.585..357H}
{Harris}, C.~R., {Millman}, K.~J., {van der Walt}, S.~J., {et~al.} 2020, \nat,
  585, 357, \dodoi{10.1038/s41586-020-2649-2}

\bibitem[{{Hawley} \& {Krolik}(2001)}]{2001ApJ...548..348H}
{Hawley}, J.~F., \& {Krolik}, J.~H. 2001, \apj, 548, 348,
  \dodoi{10.1086/318678}

\bibitem[{{Hawley} {et~al.}(2013){Hawley}, {Richers}, {Guan}, \&
  {Krolik}}]{2013ApJ...772..102H}
{Hawley}, J.~F., {Richers}, S.~A., {Guan}, X., \& {Krolik}, J.~H. 2013, \apj,
  772, 102, \dodoi{10.1088/0004-637X/772/2/102}

\bibitem[{{H{\"u}bner} {et~al.}(2022){H{\"u}bner}, {Huppenkothen}, {Lasky},
  {Inglis}, {Ick}, \& {Hogg}}]{2022ApJ...936...17H}
{H{\"u}bner}, M., {Huppenkothen}, D., {Lasky}, P.~D., {et~al.} 2022, \apj, 936,
  17, \dodoi{10.3847/1538-4357/ac7959}

\bibitem[{{Hunter}(2007)}]{2007CSE.....9...90H}
{Hunter}, J.~D. 2007, Computing in Science and Engineering, 9, 90,
  \dodoi{10.1109/MCSE.2007.55}

\bibitem[{{Hurley} {et~al.}(2013){Hurley}, {Callanan}, {Elebert}, \&
  {Reynolds}}]{2013MNRAS.430.1832H}
{Hurley}, D.~J., {Callanan}, P.~J., {Elebert}, P., \& {Reynolds}, M.~T. 2013,
  \mnras, 430, 1832, \dodoi{10.1093/mnras/stt001}

\bibitem[{{Igumenshchev}(2008)}]{2008ApJ...677..317I}
{Igumenshchev}, I.~V. 2008, \apj, 677, 317, \dodoi{10.1086/529025}

\bibitem[{{Igumenshchev} {et~al.}(2003){Igumenshchev}, {Narayan}, \&
  {Abramowicz}}]{2003ApJ...592.1042I}
{Igumenshchev}, I.~V., {Narayan}, R., \& {Abramowicz}, M.~A. 2003, \apj, 592,
  1042, \dodoi{10.1086/375769}

\bibitem[{{Inoue} {et~al.}(2017){Inoue}, {Doi}, {Tanaka}, {Sikora}, \&
  {Madejski}}]{2017ApJ...840...46I}
{Inoue}, Y., {Doi}, A., {Tanaka}, Y.~T., {Sikora}, M., \& {Madejski}, G.~M.
  2017, \apj, 840, 46, \dodoi{10.3847/1538-4357/aa6b57}

\bibitem[{{Iyer} {et~al.}(2015){Iyer}, {Nandi}, \&
  {Mandal}}]{2015ApJ...807..108I}
{Iyer}, N., {Nandi}, A., \& {Mandal}, S. 2015, \apj, 807, 108,
  \dodoi{10.1088/0004-637X/807/1/108}

\bibitem[{{Kaastra} \& {Barr}(1989)}]{1989A&A...226...59K}
{Kaastra}, J.~S., \& {Barr}, P. 1989, \aap, 226, 59

\bibitem[{{Kawaguchi} {et~al.}(1998){Kawaguchi}, {Mineshige}, {Umemura}, \&
  {Turner}}]{1998ApJ...504..671K}
{Kawaguchi}, T., {Mineshige}, S., {Umemura}, M., \& {Turner}, E.~L. 1998, \apj,
  504, 671, \dodoi{10.1086/306105}

\bibitem[{{Kelly}(2007)}]{2007ApJ...665.1489K}
{Kelly}, B.~C. 2007, \apj, 665, 1489, \dodoi{10.1086/519947}

\bibitem[{{Kelly} {et~al.}(2009){Kelly}, {Bechtold}, \&
  {Siemiginowska}}]{2009ApJ...698..895K}
{Kelly}, B.~C., {Bechtold}, J., \& {Siemiginowska}, A. 2009, \apj, 698, 895,
  \dodoi{10.1088/0004-637X/698/1/895}

\bibitem[{{Kelly} {et~al.}(2014){Kelly}, {Becker}, {Sobolewska},
  {Siemiginowska}, \& {Uttley}}]{2014ApJ...788...33K}
{Kelly}, B.~C., {Becker}, A.~C., {Sobolewska}, M., {Siemiginowska}, A., \&
  {Uttley}, P. 2014, \apj, 788, 33, \dodoi{10.1088/0004-637X/788/1/33}

\bibitem[{{Koljonen} \& {Hovatta}(2021)}]{2021A&A...647A.173K}
{Koljonen}, K.~I.~I., \& {Hovatta}, T. 2021, \aap, 647, A173,
  \dodoi{10.1051/0004-6361/202039581}

\bibitem[{{K{\"o}rding} {et~al.}(2006){K{\"o}rding}, {Falcke}, \&
  {Corbel}}]{2006A&A...456..439K}
{K{\"o}rding}, E., {Falcke}, H., \& {Corbel}, S. 2006, \aap, 456, 439,
  \dodoi{10.1051/0004-6361:20054144}

\bibitem[{{Koz{\l}owski}(2017)}]{2017A&A...597A.128K}
{Koz{\l}owski}, S. 2017, \aap, 597, A128, \dodoi{10.1051/0004-6361/201629890}

\bibitem[{{Krawczynski} {et~al.}(2022){Krawczynski}, {Muleri}, {Dov{\v{c}}iak},
  {Veledina}, {Rodriguez Cavero}, {Svoboda}, {Ingram}, {Matt}, {Garcia},
  {Loktev}, {Negro}, {Poutanen}, {Kitaguchi}, {Podgorn{\'y}}, {Rankin},
  {Zhang}, {Berdyugin}, {Berdyugina}, {Bianchi}, {Blinov}, {Capitanio}, {Di
  Lalla}, {Draghis}, {Fabiani}, {Kagitani}, {Kravtsov}, {Kiehlmann},
  {Latronico}, {Lutovinov}, {Mandarakas}, {Marin}, {Marinucci}, {Miller},
  {Mizuno}, {Molkov}, {Omodei}, {Petrucci}, {Ratheesh}, {Sakanoi}, {Semena},
  {Skalidis}, {Soffitta}, {Tennant}, {Thalhammer}, {Tombesi}, {Weisskopf},
  {Wilms}, {Zhang}, {Agudo}, {Antonelli}, {Bachetti}, {Baldini}, {Baumgartner},
  {Bellazzini}, {Bongiorno}, {Bonino}, {Brez}, {Bucciantini}, {Castellano},
  {Cavazzuti}, {Ciprini}, {Costa}, {De Rosa}, {Del Monte}, {Di Gesu}, {Di
  Marco}, {Donnarumma}, {Doroshenko}, {Ehlert}, {Enoto}, {Evangelista},
  {Ferrazzoli}, {Gunji}, {Hayashida}, {Heyl}, {Iwakiri}, {Jorstad}, {Karas},
  {Kolodziejczak}, {La Monaca}, {Liodakis}, {Maldera}, {Manfreda}, {Marscher},
  {Marshall}, {Mitsuishi}, {Ng}, {O{\textquoteright}Dell}, {Oppedisano},
  {Papitto}, {Pavlov}, {Peirson}, {Perri}, {Pesce-Rollins}, {Pilia},
  {Possenti}, {Puccetti}, {Ramsey}, {Romani}, {Sgr{\`o}}, {Slane}, {Spandre},
  {Tamagawa}, {Tavecchio}, {Taverna}, {Tawara}, {Thomas}, {Trois}, {Tsygankov},
  {Turolla}, {Vink}, {Wu}, {Xie}, \& {Zane}}]{2022Sci...378..650K}
{Krawczynski}, H., {Muleri}, F., {Dov{\v{c}}iak}, M., {et~al.} 2022, Science,
  378, 650, \dodoi{10.1126/science.add5399}

\bibitem[{{Latter} \& {Papaloizou}(2012)}]{2012MNRAS.426.1107L}
{Latter}, H.~N., \& {Papaloizou}, J.~C.~B. 2012, \mnras, 426, 1107,
  \dodoi{10.1111/j.1365-2966.2012.21748.x}

\bibitem[{{Liodakis} {et~al.}(2018){Liodakis}, {Hovatta}, {Huppenkothen},
  {Kiehlmann}, {Max-Moerbeck}, \& {Readhead}}]{2018ApJ...866..137L}
{Liodakis}, I., {Hovatta}, T., {Huppenkothen}, D., {et~al.} 2018, \apj, 866,
  137, \dodoi{10.3847/1538-4357/aae2b7}

\bibitem[{{Liodakis} {et~al.}(2017){Liodakis}, {Pavlidou}, {Papadakis},
  {Angelakis}, {Marchili}, {Zensus}, {Fuhrmann}, {Karamanavis}, {Myserlis},
  {Nestoras}, {Palaiologou}, \& {Readhead}}]{2017ApJ...851..144L}
{Liodakis}, I., {Pavlidou}, V., {Papadakis}, I., {et~al.} 2017, \apj, 851, 144,
  \dodoi{10.3847/1538-4357/aa9992}

\bibitem[{{Livio} {et~al.}(2003){Livio}, {Pringle}, \&
  {King}}]{2003ApJ...593..184L}
{Livio}, M., {Pringle}, J.~E., \& {King}, A.~R. 2003, \apj, 593, 184,
  \dodoi{10.1086/375872}

\bibitem[{{MacLeod} {et~al.}(2010){MacLeod}, {Ivezi{\'c}}, {Kochanek},
  {Koz{\l}owski}, {Kelly}, {Bullock}, {Kimball}, {Sesar}, {Westman}, {Brooks},
  {Gibson}, {Becker}, \& {de Vries}}]{2010ApJ...721.1014M}
{MacLeod}, C.~L., {Ivezi{\'c}}, {\v{Z}}., {Kochanek}, C.~S., {et~al.} 2010,
  \apj, 721, 1014, \dodoi{10.1088/0004-637X/721/2/1014}

\bibitem[{{Maraschi} \& {Tavecchio}(2003)}]{2003ApJ...593..667M}
{Maraschi}, L., \& {Tavecchio}, F. 2003, \apj, 593, 667, \dodoi{10.1086/342118}

\bibitem[{{Markoff} {et~al.}(2001){Markoff}, {Falcke}, \&
  {Fender}}]{2001A&A...372L..25M}
{Markoff}, S., {Falcke}, H., \& {Fender}, R. 2001, \aap, 372, L25,
  \dodoi{10.1051/0004-6361:20010420}

\bibitem[{{Markoff} {et~al.}(2005){Markoff}, {Nowak}, \&
  {Wilms}}]{2005ApJ...635.1203M}
{Markoff}, S., {Nowak}, M.~A., \& {Wilms}, J. 2005, \apj, 635, 1203,
  \dodoi{10.1086/497628}

\bibitem[{{McHardy} {et~al.}(2006){McHardy}, {Koerding}, {Knigge}, {Uttley}, \&
  {Fender}}]{2006Natur.444..730M}
{McHardy}, I.~M., {Koerding}, E., {Knigge}, C., {Uttley}, P., \& {Fender},
  R.~P. 2006, \nat, 444, 730, \dodoi{10.1038/nature05389}

\bibitem[{{Merloni} {et~al.}(2003){Merloni}, {Heinz}, \& {di
  Matteo}}]{2003MNRAS.345.1057M}
{Merloni}, A., {Heinz}, S., \& {di Matteo}, T. 2003, \mnras, 345, 1057,
  \dodoi{10.1046/j.1365-2966.2003.07017.x}

\bibitem[{{Mirabel} \& {Rodr{\'\i}guez}(1999)}]{1999ARA&A..37..409M}
{Mirabel}, I.~F., \& {Rodr{\'\i}guez}, L.~F. 1999, \araa, 37, 409,
  \dodoi{10.1146/annurev.astro.37.1.409}

\bibitem[{{Molla} {et~al.}(2016){Molla}, {Debnath}, {Chakrabarti}, {Mondal},
  {Jana}, \& {Chatterjee}}]{2016cosp...41E1324M}
{Molla}, A.~A., {Debnath}, D., {Chakrabarti}, S.~K., {et~al.} 2016, in 41st
  COSPAR Scientific Assembly, Vol.~41, E1.6--18--16

\bibitem[{{Moreno} {et~al.}(2019){Moreno}, {Vogeley}, {Richards}, \&
  {Yu}}]{2019PASP..131f3001M}
{Moreno}, J., {Vogeley}, M.~S., {Richards}, G.~T., \& {Yu}, W. 2019, \pasp,
  131, 063001, \dodoi{10.1088/1538-3873/ab1597}

\bibitem[{{Morgan} {et~al.}(2018){Morgan}, {Hyer}, {Bonvin}, {Mosquera},
  {Cornachione}, {Courbin}, {Kochanek}, \& {Falco}}]{2018ApJ...869..106M}
{Morgan}, C.~W., {Hyer}, G.~E., {Bonvin}, V., {et~al.} 2018, \apj, 869, 106,
  \dodoi{10.3847/1538-4357/aaed3e}

\bibitem[{{Motta} {et~al.}(2014){Motta}, {Belloni}, {Stella},
  {Mu{\~n}oz-Darias}, \& {Fender}}]{2014MNRAS.437.2554M}
{Motta}, S.~E., {Belloni}, T.~M., {Stella}, L., {Mu{\~n}oz-Darias}, T., \&
  {Fender}, R. 2014, \mnras, 437, 2554, \dodoi{10.1093/mnras/stt2068}

\bibitem[{{Mukherjee} {et~al.}(2019){Mukherjee}, {Mitra}, \&
  {Chatterjee}}]{2019MNRAS.486.1672M}
{Mukherjee}, S., {Mitra}, K., \& {Chatterjee}, R. 2019, \mnras, 486, 1672,
  \dodoi{10.1093/mnras/stz858}

\bibitem[{{Nalewajko}(2017)}]{2017Galax...5...64N}
{Nalewajko}, K. 2017, Galaxies, 5, 64, \dodoi{10.3390/galaxies5040064}

\bibitem[{{Narayan} {et~al.}(1997){Narayan}, {Barret}, \&
  {McClintock}}]{1997ApJ...482..448N}
{Narayan}, R., {Barret}, D., \& {McClintock}, J.~E. 1997, \apj, 482, 448,
  \dodoi{10.1086/304134}

\bibitem[{{Narayan} {et~al.}(2003){Narayan}, {Igumenshchev}, \&
  {Abramowicz}}]{2003PASJ...55L..69N}
{Narayan}, R., {Igumenshchev}, I.~V., \& {Abramowicz}, M.~A. 2003, \pasj, 55,
  L69, \dodoi{10.1093/pasj/55.6.L69}

\bibitem[{{Narayan} \& {Yi}(1994)}]{1994ApJ...428L..13N}
{Narayan}, R., \& {Yi}, I. 1994, \apjl, 428, L13, \dodoi{10.1086/187381}

\bibitem[{{Nisbet} \& {Best}(2016)}]{2016MNRAS.455.2551N}
{Nisbet}, D.~M., \& {Best}, P.~N. 2016, \mnras, 455, 2551,
  \dodoi{10.1093/mnras/stv2450}

\bibitem[{{Orosz} {et~al.}(2011){Orosz}, {McClintock}, {Aufdenberg},
  {Remillard}, {Reid}, {Narayan}, \& {Gou}}]{2011ApJ...742...84O}
{Orosz}, J.~A., {McClintock}, J.~E., {Aufdenberg}, J.~P., {et~al.} 2011, \apj,
  742, 84, \dodoi{10.1088/0004-637X/742/2/84}

\bibitem[{{Padovani} {et~al.}(2017){Padovani}, {Alexander}, {Assef}, {De
  Marco}, {Giommi}, {Hickox}, {Richards}, {Smol{\v{c}}i{\'c}},
  {Hatziminaoglou}, {Mainieri}, \& {Salvato}}]{2017A&ARv..25....2P}
{Padovani}, P., {Alexander}, D.~M., {Assef}, R.~J., {et~al.} 2017, \aapr, 25,
  2, \dodoi{10.1007/s00159-017-0102-9}

\bibitem[{{Paliya} {et~al.}(2021){Paliya}, {Dom{\'\i}nguez}, {Ajello},
  {Olmo-Garc{\'\i}a}, \& {Hartmann}}]{2021ApJS..253...46P}
{Paliya}, V.~S., {Dom{\'\i}nguez}, A., {Ajello}, M., {Olmo-Garc{\'\i}a}, A., \&
  {Hartmann}, D. 2021, \apjs, 253, 46, \dodoi{10.3847/1538-4365/abe135}

\bibitem[{{Picchi} {et~al.}(2020){Picchi}, {Shore}, {Harvey}, \&
  {Berdyugin}}]{2020AA...640A..96P}
{Picchi}, P., {Shore}, S.~N., {Harvey}, E.~J., \& {Berdyugin}, A. 2020, \aap,
  640, A96, \dodoi{10.1051/0004-6361/202037960}

\bibitem[{{Remillard} \& {McClintock}(2006)}]{2006ARAA..44...49R}
{Remillard}, R.~A., \& {McClintock}, J.~E. 2006, \araa, 44, 49,
  \dodoi{10.1146/annurev.astro.44.051905.092532}

\bibitem[{{Ricci} {et~al.}(2022){Ricci}, {Boccardi}, {Nokhrina}, {Perucho},
  {MacDonald}, {Mattia}, {Grandi}, {Madika}, {Krichbaum}, \&
  {Zensus}}]{2022A&A...664A.166R}
{Ricci}, L., {Boccardi}, B., {Nokhrina}, E., {et~al.} 2022, \aap, 664, A166,
  \dodoi{10.1051/0004-6361/202243958}

\bibitem[{{Rodriguez} {et~al.}(2011){Rodriguez}, {Corbel}, {Caballero},
  {Tomsick}, {Tzioumis}, {Paizis}, {Cadolle Bel}, \&
  {Kuulkers}}]{2011A&A...533L...4R}
{Rodriguez}, J., {Corbel}, S., {Caballero}, I., {et~al.} 2011, \aap, 533, L4,
  \dodoi{10.1051/0004-6361/201117511}

\bibitem[{{Ruan} {et~al.}(2012){Ruan}, {Anderson}, {MacLeod}, {Becker},
  {Burnett}, {Davenport}, {Ivezi{\'c}}, {Kochanek}, {Plotkin}, {Sesar}, \&
  {Stuart}}]{2012ApJ...760...51R}
{Ruan}, J.~J., {Anderson}, S.~F., {MacLeod}, C.~L., {et~al.} 2012, \apj, 760,
  51, \dodoi{10.1088/0004-637X/760/1/51}

\bibitem[{{Ryan} {et~al.}(2019){Ryan}, {Siemiginowska}, {Sobolewska}, \&
  {Grindlay}}]{2019ApJ...885...12R}
{Ryan}, J.~L., {Siemiginowska}, A., {Sobolewska}, M.~A., \& {Grindlay}, J.
  2019, \apj, 885, 12, \dodoi{10.3847/1538-4357/ab426a}

\bibitem[{{Scaringi} {et~al.}(2015){Scaringi}, {Maccarone}, {Kording},
  {Knigge}, {Vaughan}, {Marsh}, {Aranzana}, {Dhillon}, \&
  {Barros}}]{2015SciA....1E0686S}
{Scaringi}, S., {Maccarone}, T.~J., {Kording}, E., {et~al.} 2015, Science
  Advances, 1, e1500686, \dodoi{10.1126/sciadv.1500686}

\bibitem[{{Shakura} \& {Sunyaev}(1973)}]{1973A&A....24..337S}
{Shakura}, N.~I., \& {Sunyaev}, R.~A. 1973, \aap, 24, 337

\bibitem[{{Sharma} {et~al.}(2024){Sharma}, {Kamaram}, {Prince}, {Khatoon}, \&
  {Bose}}]{2024MNRAS.527.2672S}
{Sharma}, A., {Kamaram}, S.~R., {Prince}, R., {Khatoon}, R., \& {Bose}, D.
  2024, \mnras, 527, 2672, \dodoi{10.1093/mnras/stad3399}

\bibitem[{{Smak}(1999)}]{1999AcA....49..391S}
{Smak}, J. 1999, \actaa, 49, 391

\bibitem[{{Str{\"u}der} {et~al.}(2001){Str{\"u}der}, {Briel}, {Dennerl},
  {Hartmann}, {Kendziorra}, {Meidinger}, {Pfeffermann}, {Reppin}, {Aschenbach},
  {Bornemann}, {Br{\"a}uninger}, {Burkert}, {Elender}, {Freyberg}, {Haberl},
  {Hartner}, {Heuschmann}, {Hippmann}, {Kastelic}, {Kemmer}, {Kettenring},
  {Kink}, {Krause}, {M{\"u}ller}, {Oppitz}, {Pietsch}, {Popp}, {Predehl},
  {Read}, {Stephan}, {St{\"o}tter}, {Tr{\"u}mper}, {Holl}, {Kemmer}, {Soltau},
  {St{\"o}tter}, {Weber}, {Weichert}, {von Zanthier}, {Carathanassis}, {Lutz},
  {Richter}, {Solc}, {B{\"o}ttcher}, {Kuster}, {Staubert}, {Abbey}, {Holland},
  {Turner}, {Balasini}, {Bignami}, {La Palombara}, {Villa}, {Buttler},
  {Gianini}, {Lain{\'e}}, {Lumb}, \& {Dhez}}]{2001A&A...365L..18S}
{Str{\"u}der}, L., {Briel}, U., {Dennerl}, K., {et~al.} 2001, \aap, 365, L18,
  \dodoi{10.1051/0004-6361:20000066}

\bibitem[{{Suberlak} {et~al.}(2021){Suberlak}, {Ivezi{\'c}}, \&
  {MacLeod}}]{2021ApJ...907...96S}
{Suberlak}, K.~L., {Ivezi{\'c}}, {\v{Z}}., \& {MacLeod}, C. 2021, \apj, 907,
  96, \dodoi{10.3847/1538-4357/abc698}

\bibitem[{{Sun} {et~al.}(2020){Sun}, {Xue}, {Brandt}, {Gu}, {Trump}, {Cai},
  {He}, {Lin}, {Liu}, \& {Wang}}]{2020ApJ...891..178S}
{Sun}, M., {Xue}, Y., {Brandt}, W.~N., {et~al.} 2020, \apj, 891, 178,
  \dodoi{10.3847/1538-4357/ab789e}

\bibitem[{{Takahashi} {et~al.}(2009){Takahashi}, {Kishishita}, {Uchiyama},
  {Tanaka}, {Yamaoka}, {Khangulyan}, {Aharonian}, {Bosch-Ramon}, \&
  {Hinton}}]{2009ApJ...697..592T}
{Takahashi}, T., {Kishishita}, T., {Uchiyama}, Y., {et~al.} 2009, \apj, 697,
  592, \dodoi{10.1088/0004-637X/697/1/592}

\bibitem[{{Tian} {et~al.}(2023){Tian}, {Zhang}, {Wang}, {Wang}, {Sun}, {Liu},
  {Zhang}, {Dai}, {Yuan}, {Zhang}, {Liu}, {Jiang}, {Wu}, {Zheng}, {Chen}, {Li},
  {Zhu}, {Pan}, {Gan}, {Chen}, \& {Sai}}]{2023Natur.621..271T}
{Tian}, P., {Zhang}, P., {Wang}, W., {et~al.} 2023, \nat, 621, 271,
  \dodoi{10.1038/s41586-023-06336-6}

\bibitem[{{Urry}(1996)}]{1996ASPC..110..391U}
{Urry}, C.~M. 1996, in Astronomical Society of the Pacific Conference Series,
  Vol. 110, Blazar Continuum Variability, ed. H.~R. {Miller}, J.~R. {Webb}, \&
  J.~C. {Noble}, 391, \dodoi{10.48550/arXiv.astro-ph/9609023}

\bibitem[{{Urry} \& {Padovani}(1995)}]{1995PASP..107..803U}
{Urry}, C.~M., \& {Padovani}, P. 1995, \pasp, 107, 803, \dodoi{10.1086/133630}

\bibitem[{{Vila}(2012)}]{2012BAAA...55..539V}
{Vila}, G.~S. 2012, Boletin de la Asociacion Argentina de Astronomia La Plata
  Argentina, 55, 539

\bibitem[{{Wang} {et~al.}(2021){Wang}, {Mastroserio}, {Kara}, {Garc{\'\i}a},
  {Ingram}, {Connors}, {van der Klis}, {Dauser}, {Steiner}, {Buisson}, {Homan},
  {Lucchini}, {Fabian}, {Bright}, {Fender}, {Cackett}, \&
  {Remillard}}]{2021ApJ...910L...3W}
{Wang}, J., {Mastroserio}, G., {Kara}, E., {et~al.} 2021, \apjl, 910, L3,
  \dodoi{10.3847/2041-8213/abec79}

\bibitem[{{Wang} {et~al.}(2004){Wang}, {Luo}, \& {Ho}}]{2004ApJ...615L...9W}
{Wang}, J.-M., {Luo}, B., \& {Ho}, L.~C. 2004, \apjl, 615, L9,
  \dodoi{10.1086/426060}

\bibitem[{{Williams}(1995)}]{1995PhRvD..51.5387W}
{Williams}, R.~K. 1995, \prd, 51, 5387, \dodoi{10.1103/PhysRevD.51.5387}

\bibitem[{{Williams}(2004)}]{2004ApJ...611..952W}
---. 2004, \apj, 611, 952, \dodoi{10.1086/422304}

\bibitem[{{Wilms} {et~al.}(2006){Wilms}, {Nowak}, {Pottschmidt}, {Pooley}, \&
  {Fritz}}]{2006AA...447..245W}
{Wilms}, J., {Nowak}, M.~A., {Pottschmidt}, K., {Pooley}, G.~G., \& {Fritz}, S.
  2006, \aap, 447, 245, \dodoi{10.1051/0004-6361:20053938}

\bibitem[{{Xie} {et~al.}(2020){Xie}, {Yan}, \& {Wu}}]{2020ApJ...891...31X}
{Xie}, F.-G., {Yan}, Z., \& {Wu}, Z. 2020, \apj, 891, 31,
  \dodoi{10.3847/1538-4357/ab711f}

\bibitem[{{Yang} {et~al.}(2021){Yang}, {Yan}, {Zhang}, {Dai}, \&
  {Zhang}}]{2021ApJ...907..105Y}
{Yang}, S., {Yan}, D., {Zhang}, P., {Dai}, B., \& {Zhang}, L. 2021, \apj, 907,
  105, \dodoi{10.3847/1538-4357/abcbff}

\bibitem[{{You} {et~al.}(2023){You}, {Cao}, {Yan}, {Hameury}, {Czerny}, {Wu},
  {Xia}, {Sikora}, {Zhang}, {Du}, \& {Zycki}}]{2023Sci...381..961Y}
{You}, B., {Cao}, X., {Yan}, Z., {et~al.} 2023, Science, 381, 961,
  \dodoi{10.1126/science.abo4504}

\bibitem[{{Yuan} \& {Narayan}(2014)}]{2014ARA&A..52..529Y}
{Yuan}, F., \& {Narayan}, R. 2014, \araa, 52, 529,
  \dodoi{10.1146/annurev-astro-082812-141003}

\bibitem[{{Zabalza} {et~al.}(2011){Zabalza}, {Paredes}, \&
  {Bosch-Ramon}}]{2011AA...527A...9Z}
{Zabalza}, V., {Paredes}, J.~M., \& {Bosch-Ramon}, V. 2011, \aap, 527, A9,
  \dodoi{10.1051/0004-6361/201015373}

\bibitem[{{Zamaninasab} {et~al.}(2014){Zamaninasab}, {Clausen-Brown},
  {Savolainen}, \& {Tchekhovskoy}}]{2014Natur.510..126Z}
{Zamaninasab}, M., {Clausen-Brown}, E., {Savolainen}, T., \& {Tchekhovskoy}, A.
  2014, \nat, 510, 126, \dodoi{10.1038/nature13399}

\bibitem[{{Zdziarski} {et~al.}(2002){Zdziarski}, {Poutanen}, {Paciesas}, \&
  {Wen}}]{2002ApJ...578..357Z}
{Zdziarski}, A.~A., {Poutanen}, J., {Paciesas}, W.~S., \& {Wen}, L. 2002, \apj,
  578, 357, \dodoi{10.1086/342402}

\bibitem[{{Zdziarski} {et~al.}(2020){Zdziarski}, {Shapopi}, \&
  {Pooley}}]{2020ApJ...894L..18Z}
{Zdziarski}, A.~A., {Shapopi}, J.~N.~S., \& {Pooley}, G.~G. 2020, \apjl, 894,
  L18, \dodoi{10.3847/2041-8213/ab8d3b}

\bibitem[{{Zhang} {et~al.}(2022){Zhang}, {Yan}, \&
  {Zhang}}]{2022ApJ...930..157Z}
{Zhang}, H., {Yan}, D., \& {Zhang}, L. 2022, \apj, 930, 157,
  \dodoi{10.3847/1538-4357/ac679e}

\bibitem[{{Zhang} {et~al.}(2023{\natexlab{a}}){Zhang}, {Yan}, \&
  {Zhang}}]{2023ApJ...944..103Z}
---. 2023{\natexlab{a}}, \apj, 944, 103, \dodoi{10.3847/1538-4357/acafe5}

\bibitem[{{Zhang} {et~al.}(2023{\natexlab{b}}){Zhang}, {Yang}, \&
  {Dai}}]{2023ApJ...946...52Z}
{Zhang}, H., {Yang}, S., \& {Dai}, B. 2023{\natexlab{b}}, \apj, 946, 52,
  \dodoi{10.3847/1538-4357/acbe37}

\bibitem[{{Zhang} {et~al.}(2012){Zhang}, {Liang}, {Zhang}, \&
  {Bai}}]{2012ApJ...752..157Z}
{Zhang}, J., {Liang}, E.-W., {Zhang}, S.-N., \& {Bai}, J.~M. 2012, \apj, 752,
  157, \dodoi{10.1088/0004-637X/752/2/157}

\bibitem[{{Zu} {et~al.}(2013){Zu}, {Kochanek}, {Koz{\l}owski}, \&
  {Udalski}}]{2013ApJ...765..106Z}
{Zu}, Y., {Kochanek}, C.~S., {Koz{\l}owski}, S., \& {Udalski}, A. 2013, \apj,
  765, 106, \dodoi{10.1088/0004-637X/765/2/106}

\end{thebibliography}
\bibliographystyle{aasjournal}	
\end{document}